\documentclass[aps,
 ,twocolumn
,superscriptaddress,
]{revtex4-1}

\usepackage{mathptmx}
\DeclareMathAlphabet{\mathcal}{OMS}{cmsy}{m}{n}

\usepackage{graphicx}
\usepackage{amsmath}
\usepackage{enumitem}
\usepackage{amssymb}
\usepackage{array}
\newcolumntype{L}[1]{>{\raggedright\let\newline\\\arraybackslash\hspace{0pt}}m{#1}}
\newcolumntype{C}[1]{>{\centering\let\newline\\\arraybackslash\hspace{0pt}}m{#1}}
\newcolumntype{R}[1]{>{\raggedleft\let\newline\\\arraybackslash\hspace{0pt}}m{#1}}
\usepackage{mathrsfs}
\usepackage{dsfont}
\usepackage{appendix}
\usepackage{xtab,afterpage}
\usepackage{hyperref}
\usepackage{multirow}
\usepackage{tikz}
\usepackage{pgflibraryarrows}
\usepackage{pgflibrarysnakes}
\usetikzlibrary{decorations.pathmorphing}
\usepgflibrary{arrows.meta}

\usepackage{subfigure}

\newcommand{\D}{\mathrm{d}}

\newcommand{\dd}{\dagger}

\newcommand{\langlem}{\langle 0_M|} 
\newcommand{\ranglem}{|0_M\rangle}
\newcommand{\omgea}{\omega}

\usepackage{xtab,afterpage}
\usepackage{hyperref}
\usepackage{braket}

\graphicspath{{Figures/}}
\allowdisplaybreaks

\definecolor{blueblue}{RGB}{21,47,181}

\hypersetup{ linktoc=all,
    colorlinks, linkcolor={blueblue},
    citecolor={blueblue}, urlcolor={blueblue}
}

\begin{document}

\title{Generating multi-partite entanglement from the quantum vacuum with a finite-lifetime mirror}

\author{Joshua Foo}
\email{joshua.foo@uqconnect.edu.au}
\affiliation{Centre for Quantum Computation and Communication Technology, School of Mathematics and Physics, The University of Queensland, St. Lucia, Queensland, 4072, Australia}
\author{Sho Onoe}
\affiliation{Centre for Quantum Computation and Communication Technology, School of Mathematics and Physics, The University of Queensland, St. Lucia, Queensland, 4072, Australia}
\author{Magdalena Zych}
\affiliation{Centre for Engineered Quantum Systems, School of Mathematics and Physics, The University of Queensland, St. Lucia, Queensland, 4072, Australia}
\author{Timothy.~C.~Ralph}
\affiliation{Centre for Quantum Computation and Communication Technology, School of Mathematics and Physics, The University of Queensland, St. Lucia, Queensland, 4072, Australia}
\email{ralph@physics.uq.edu.au}

\date{\today}

\begin{abstract}
{Observers following special classes of finite-lifetime trajectories have been shown to experience an effective temperature, a generalisation of the Unruh temperature for uniformly accelerated observers. We consider a mirror following such a trajectory -- and is hence localised to a strictly bounded causal diamond -- that perfectly reflects incoming field modes. We find that inertial observers in the Minkowski vacuum detect particles along the half null-rays at the beginning and end of the mirror's lifetime. These particle distributions exhibit multi-partite entanglement, which reveals novel structure within the vacuum correlations. The interaction is modelled using a non-perturbative circuit model and does not suffer from energy divergences.}
\end{abstract}
\maketitle

\textit{Introduction}\textemdash Using the `thermal-time hypothesis', Martinetti and Rovelli have demonstrated the existence of `Unruh-like' effects (namely, a thermalised vacuum state) for observers localised to bounded regions of spacetime \cite{connes1994neumann,martinetti2003diamond}. These regions are known as spacetime or causal diamonds, and are formed by overlapping the future lightcone at the observer's birth with the past lightcone at their death. For the special case of a stationary diamond observer, the temperature they experience is given by 
\begin{align}
    T_D &= \frac{2}{\pi \mathcal{T}}
\end{align}
where $\mathcal{T}$ is their lifetime \cite{martinetti2003diamond}. The origin of this effect lies in the entanglement structure of the Minkowski vacuum, which can be expressed as a two-mode squeezed state of the modes with support inside the diamond, and those external to it \cite{birrelldavies}. Since the observer only has access to the interior modes in their restricted section of spacetime, tracing out the unobserved modes yields a thermal state in their frame of reference \cite{unruh1976notes,crispino2008unruh,alsing2004simplified,su2016spacetime}. More recently, a physical interpretation was given for this phenomenon, whereby a finite-lifetime Unruh-deWitt detector was shown to register the same thermal response \cite{su2016spacetime}. The detector considered is a two-level system with a particular time-dependent energy gap which tends to infinity at the beginning and end of its lifetime \cite{su2016spacetime}. 

In relativistic quantum field theory (QFT), free particles are \textit{delocalised in spacetime}, usually constructed as superpositions of complex-valued, positive-frequency field modes \cite{schlieder1965some,haag1965does,knight1961strict}. Notably, an observer confined to a causal diamond perceives and interacts with field modes which are \textit{strictly localised in spacetime}. This localisation arises naturally within the reference frame of the diamond observer, and is achieved without imposing sharp switching functions upon the interactions, which can lead to spurious dynamical effects \cite{satz2007then,louko2006often}. Moreover, localisation via causal diamonds contrasts other methods of confining quantum fields to finite regions such as cavity QFT, which imposes physical boundaries on the modes \cite{friis2012quantum,bruschi2012voyage,brown2015does,brown2015smooth}. Hence, we anticipate that diamond observers and the modes they detect and interact with may engender novel approaches to problems where localisation is important, such as the study of causal structure \cite{oreshkov2012quantum}. 

\begin{figure*}[t]
    \centering
    \includegraphics[width=0.75\linewidth]{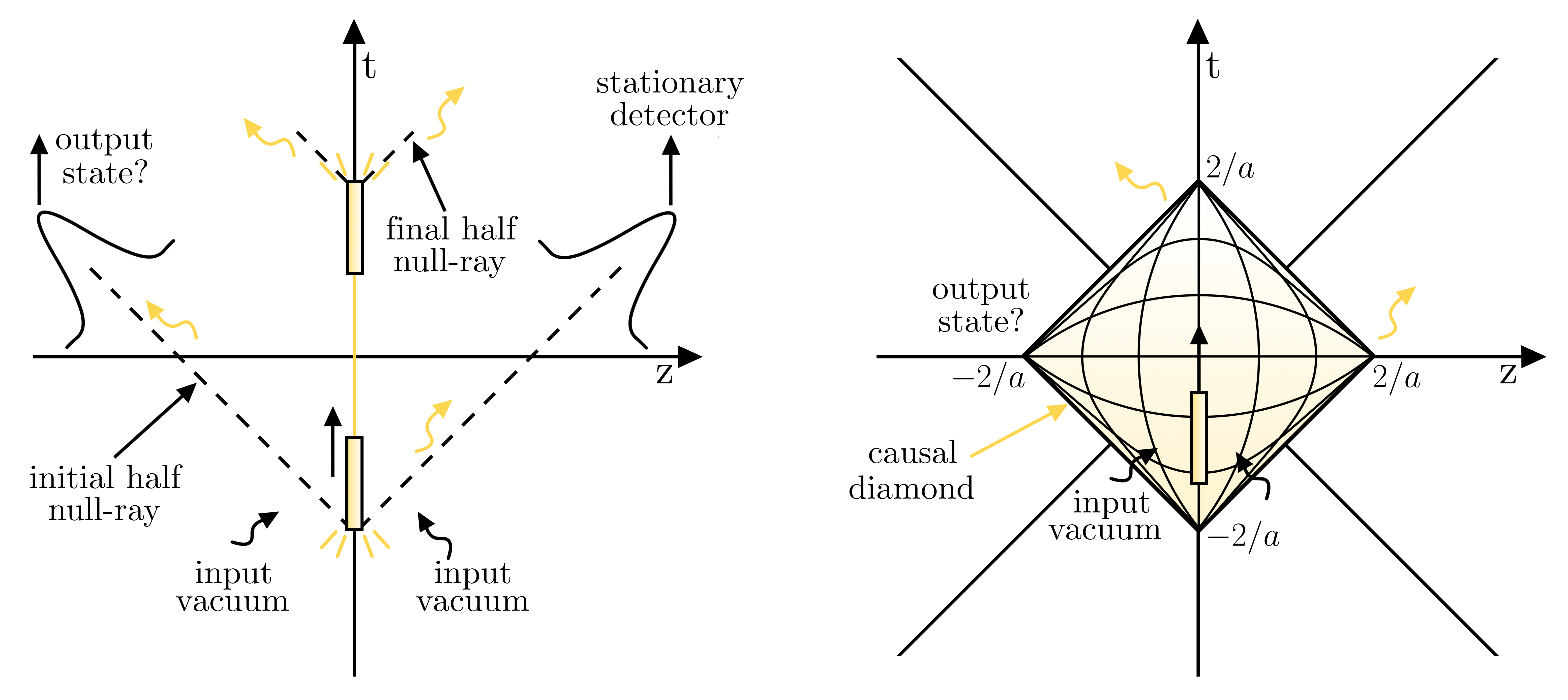}
    \caption{Schematic diagram of the finite-lifetime mirror (left) and the mirror strictly localised within a causal diamond (right). Incoming vacua are reflected off the mirror and are detected by inertial Minkowski detectors on either side of the mirror. In the diamond, the lines of latitude and longitude represent constant $\eta, \xi$ respectively.}
    \label{fig:schematic}
\end{figure*}

In this paper, we consider a finite-lifetime observer (henceforth, interchangeable with diamond observer) who interacts with strictly localised field modes using a mirror. A well-known result in relativistic QFT is that perfectly reflecting mirrors act as boundaries which affect the vacuum state of the field \cite{birrelldavies,fulling1976radiation}. For certain accelerated trajectories, the radiation flux has a thermal character, analogous to the Hawking radiation emitted from a black hole formed via gravitational collapse \cite{walker1985particle,carlitz1987reflections}. By modelling the interaction with the mirror using the non-perturbative quantum circuit framework introduced in \cite{su2017quantum,su2019decoherence}, we find that inertial observers in the Minkowski vacuum detect particle production along the initial and final half null-rays of the finite-lifetime observer's diamond (see Fig.\ \ref{fig:schematic}). Our main result demonstrates the existence of genuine, multi-partite entanglement between the outgoing left- and right-moving modes. We argue that this observation would be a distinct signature of vacuum entanglement. The amount of entanglement is maximised when the Minkowski detectors are spatially unresolved, and would be observable given current experimental limits. 

This paper is organised as follows: we first describe the coordinate system of a diamond observer and the Bogoliubov transformations between the inertial and diamond reference frame. Using these, we derive output Minkowski operators after the interaction with the finite-lifetime mirror, and demonstrate the production of particles along the initial and final half null-rays of the diamond, from the point of view of inertial observers. We present numerical results showing multi-partite entanglement between these particles, before considering the optimum case. 
Finally, we show that the non-perturbative interaction model that we utilise \cite{su2019decoherence} circumvents the usual energy divergences of the (1+1)-dimensional massless scalar field \cite{coleman1973there}, before offering some conclusions. In this paper, we use natural units $c = \hslash = k_B = 1$. 

\textit{Coordinates, field modes, and Bogoliubov transformations}\textemdash 
A static observer (who stays at $\textbf{r}=(x,y,z)= 0$) with a finite lifetime lives in a causal diamond, defined as the overlapping region between the future and past lightcones at their birth and death respectively. The diamond region satisfies $|t| + |\textbf{r}| < 2/a$ where $\mathcal{T} = 4/a$ is the lifetime of the observer. To describe spacetime events and field modes within the diamond, we introduce diamond coordinates ($\eta, \xi,\zeta,\rho)$ which are related to the Minkowski coordinates via the transformation \cite{su2016spacetime}
\begin{align}
    \eta &= a^{-1} \tanh^{-1} \bigg\{ \frac{at}{1 + (at/2)^2 - (ar/2)^2} \bigg\} \vphantom{\frac{\sqrt{1^2}}{1^2}} \\
    \xi &= a^{-1} \ln \bigg\{ \frac{\sqrt{(1 + (at/2)^2 - (ar/2)^2 )^2 - (at)^2}}{f(t,x,y,z;a)} \bigg\} \\
    \zeta &= \frac{2y}{f(t,x,y,z;a)}  \vphantom{\frac{\sqrt{1^2}}{1^2}}\\
    \rho &= \frac{2z}{f(t,x,y,z;a)} \vphantom{\frac{\sqrt{1^2}}{1^2}}
\end{align}
where $f(t,x,y,z;a) = 1 - (at/2)^2 +(ar/2)^2 -ax$ and $r = \sqrt{x^2 + y^2 + z^2}$. Intriguingly, there exists a conformal transformation that maps the bounded diamond region to the right Rindler wedge (unbounded), and an analogous transformation maps the region outside the diamond to the left Rindler wedge \cite{martinetti2003diamond}. This property provides some physical intuition to understand correlations between field modes inside and outside the diamond. We also note that recent works have drawn connections between the diamond region to the spacetime associated with the interior and exterior horizons of a Reissner-Nordstr\"om black hole \cite{de2018light,de2019light}, as well as the static patch of de Sitter spacetime \cite{tian2005sitter,good2020mirror}. 

For the remainder of this paper, we consider a static diamond observer in (1+1)-dimensional Minkowski spacetime who uses the following coordinates to describe events and distances in their reference frame,
\begin{align}
    \eta &= a^{-1} \tanh^{-1} \bigg\{ \frac{at}{1 +(at/2)^2 } \bigg\} \\
    \xi &= 0 .
\end{align}
Along the observer's worldline, $t = 2a^{-1} \tanh(a\eta/2)$ or $\D t = \D \eta/\cosh^2(a\eta/2)$. At $\eta=0$, their clock ticks at the same rate as that of an inertial Minkowski observer, whilst at $\eta = \pm \infty$, it ticks at an infinite rate. The observer interacts with a massless scalar field $\hat{\Phi}$, for which the Klein-Gordon equation $\Box\hat{\Phi}=0$ admits plane wave solutions of the form,
\begin{align}\label{eqn:minkowskifield}
    \hat{\Phi} (U,V) &= \int \D k ( \hat{a}_{kl} u_k(V) + \hat{a}_{kr} u_k(U)  + \text{H.c} \big),
\end{align}
where H.c denotes the Hermitian conjugate, $\hat{a}_{kl(r)}$ are Minkowski annihilation operators corresponding to the left-moving, $u_k(V) = (4\pi k )^{-1/2} \exp(-ikV)$, and right-moving, $u_k(U) = (4\pi k )^{-1/2}\exp(-ikU)$, positive-frequency mode functions with frequency $k$, and $V = t + z$, $U = t - z$ are lightcone coordinates  \cite{birrelldavies}. In (1+1)-dimensional spacetime, the left- and right-moving modes are decoupled, so for simplicity we treat only the left-moving sector of the field until noted otherwise. Analogous results hold for the right-moving sector.

One can also expand the field $\hat{\Phi}(V)$ in terms of the single-frequency diamond mode functions \cite{su2016spacetime}, 
\begin{align}\label{mink}
    g_\omega^{(0)} (V) &= \frac{1}{\sqrt{4\pi\omega}}  \left( \frac{1 + aV/2}{1 - aV/2} \right)^{-i\omega/a} \theta(2/a - |V|) \\
    g_\omega^\text{(ex)}(V) &= \frac{1}{\sqrt{4\pi\omega}} \left( \frac{aV/2 + 1}{aV/2 - 1} \right)^{i\omega/a} \theta(|V|- 2/a) 
\end{align}
which have support inside and outside the diamond respectively. That is, 
\begin{align}
    \hat{\Phi}(V) &= \int\D \omega \big( \hat{b}_{\omega l }g_\omega^{(0)}(V) + \hat{b}_{\omega l}g_\omega^\text{(ex)}(V) + \text{h.c} \big)
\end{align}
where $\hat{b}_{\omega l}^{(0)}$ and $\hat{b}_{\omega l}^\text{(ex)}$ are the bosonic operators which have support inside and external to the diamond. The Bogoliubov transformation between the diamond and Minkowski mode operators is given by \cite{su2016spacetime}
\begin{align}
    \hat{b}_{\omega l}^{(0)} &= \int \D k \big( \alpha_{\omega k}^{(0)} \hat{a}_{kl} + \beta^{(0)}_{\omega k}\hat{a}_{kl}^\dd \big) .
\end{align}
By directly calculating the non-zero Bogoliubov coefficients via the Klein-Gordon inner product, 
\begin{align}
    \alpha_{\omega k}^{(0)} &= \langle g_\omega^{(0)}(V), u_k(V) \rangle \nonumber \\ \label{alpha}
    &= \frac{2}{a}\frac{\sqrt{\Omega\kappa}}{\sinh(\pi\Omega)}e^{2i\kappa} M(1+i\Omega, 2, - 4i\kappa)  \\ \label{beta}
    \beta_{\omega k}^{(0)} &= \langle g_\omega^{(0)} (V), u_k^\star(V) \rangle \nonumber \\
    &= - \frac{2}{a} \frac{\sqrt{\Omega\kappa}}{\sinh(\pi\Omega)}e^{2i\kappa}M(1-i\Omega, 2, -4i\kappa) 
\end{align}
where $M(a,b,z)$ is the first-order confluent hypergeometric function \cite{abramowitz1999ia} and $\kappa = k/a, \Omega = \omega/a$, one clearly finds that the two quantisations are inequivalent; the Minkowski vacuum is not a vacuum inside the diamond, and vice versa \cite{su2016spacetime}. In particular, the particle number inside the diamond,
\begin{align}
    \langlem \hat{b}_{\omega l}^{(0)\dd} \hat{b}_{\omega'l}^{(0)}\ranglem &= \int\D k \: \beta_{\omega k}^{(0)\star} \beta_{\omega'k}^{(0)} = \frac{\delta(\omega- \omgea')}{e^{2\pi\omega/a}-1}
\end{align}
gives the familiar Planck spectrum with temperature $a/2\pi = 2/\pi\mathcal{T}$. In \cite{su2016spacetime}, it is shown that an Unruh-deWitt detector whose energy gap scales as $(1-a^2t^2/4)^{-1}$ responds to the Minkowski vacuum as if it were radiating thermal particles at this temperature. 

Before calculating the output state after the interaction, it is useful to define a set of modes that span the entirety of Minkowski spacetime from the diamond operators. We define Unruh operators, $\hat{c}_\omega^{(0)}, \hat{c}_\omega^\text{(ex)}$, that share a vacuum with the Minkowski operators, and are given by
\begin{align}\label{unruh1}
    \hat{c}_{\omega i}^{(0)} &= \cosh r_\omega \hat{b}_{\omega i}^{(0)} - \sinh r_\omega \hat{b}_{\omega i}^\text{(ex)$\dd$} \\ \label{unruh2}
    \hat{c}_{\omega i}^\text{(ex)} &= \cosh r_\omega \hat{b}_{\omega i}^\text{(ex)} - \sinh r_\omega \hat{b}_{\omega i}^{(0)\dd}
\end{align}
where $r_\omega = \tanh^{-1}\exp(-\pi\Omega)$, $i = l,r$, and $\hat{b}_{\omega i}^\text{(ex)}$ is the annihilation operator associated with the external modes, which can be similarly decomposed as a linear combination of Minkowski operators, as per Eq.\ (\ref{mink}). The inverse transformation is given by,
\begin{align}\label{vacua}
    \hat{b}_{\omega i}^{(0)} &= \cosh r_\omega \hat{c}_{\omega i}^{(0)} + \sinh r_\omega \hat{c}_\omega^\text{(ex)$\dd$} \\
    \hat{b}_{\omega i}^\text{(ex)} &= \cosh r_\omega \hat{c}_{\omega i}^\text{(ex)} + \sinh r_\omega \hat{c}_{\omega i}^{(0)\dd}.
\end{align}
Eq.\ (\ref{vacua}) describes the state inside the diamond as a two-mode squeezed state of the Minkowski vacuum. 

\textit{Interaction model}\textemdash To model interactions between the finite-lifetime observer and incoming field modes, we use the non-perturbative quantum circuit approach developed recently \cite{su2019decoherence,su2017quantum}. In the Heisenberg picture, input Unruh operators $\hat{c}_{\omega i}^{(0)}, \hat{c}_{\omega i}^\text{(ex)}$ are transformed into the diamond operators $\hat{b}_{\omega i}^{(0)},\hat{b}_{\omega i}^\text{(ex)}$. In particular, $\hat{b}_\omega^{(0)}$ are the natural modes which the diamond observer interacts with. The input diamond operators are evolved unitarily into output diamond operators (in our case, perfectly reflected off the mirror), and then transformed into output Unruh operators. These are used to construct output Minkowski operators which inertial observers can detect. Inside the diamond, we consider time-dependent interactions with wavepacket diamond modes,
\begin{align}
    \hat{b}_{gi}^{(0)} &= \int \D \omega\: g(\omega) \hat{b}_{\omega i}^{(0)},
\end{align}
where $g(\omega)$ takes the form
\begin{align}
    g(\omega) &= \mathcal{N}_\omega\sqrt{\mathcal{\omega}} \exp \left[ - \frac{(\omega - \omega_0)^2}{4\delta^2} \right]
\end{align}
and $\mathcal{N}_\omega $ ensures that $\int\D \omega\:|g(\omgea)|^2 = 1$. In the narrowband, high-frequency limit ($\omega_0\gg \delta)$, $g(\omega)$ has an approximately Gaussian profile whereby $\omega_0$ is the centre-frequency and $\delta$ the bandwidth of the mode. Moreover, $g(\omega)$ can be understood in the time-domain as a Gaussian pulse centred at $t = 0$. 

In the following, unprimed operators are associated with input modes prior to the interaction while primed operators correspond to output modes after the interaction. The input-output relation between the incoming diamond operator $\hat{b}_{\omega i}^{(0)}$ and the outgoing operator $\hat{b}_{\omega i}^{(0)\prime}$ is given by \cite{su2019decoherence,rohde2007spectral}
\begin{align}\label{eqn:Rindlerwavepacketeqn}
    \hat{b}_{\omega i}^{(0)\prime} &= \hat{b}_{\omega i}^{(0)} + g^\star(\omega) \big( \hat{U}_g^\dd \hat{b}_{gi}^{(0)} \hat{U}_g - \hat{b}_{gi}^{(0)} \big),
\end{align}
where $\hat{U}_g$ is a general unitary transformation enacted by the diamond observer. Since the diamond observer does not have access to the external modes, these remain unaffected by the unitary, $\hat{b}_{\omega i}^\text{(ex)$\prime$} = \hat{b}_{\omega i}^\text{(ex)}$. For the finite-lifetime mirror, the interaction $\hat{U}_g$ can be modelled as a beamsplitter transformation \cite{weedbrook2012gaussian},
\begin{align} \label{UG}
    \hat{U}_g = \exp \left[ - i\theta \big( e^{i\phi}\hat{b}_{gl}^{(0)\dd} \hat{b}_{gr}^{(0)} + e^{-i\phi} \hat{b}_{gl}^{(0)}\hat{b}_{gr}^{(0)\dd } \big) \right]
\end{align}
which transforms the incoming wavepacket modes $\hat{b}_{gl}^{(0)}$, $\hat{b}_{gr}^{(0)}$ into the output modes $\hat{b}_{gl}^{(0)\prime},\hat{b}_{gr}^{(0)\prime}$ as
\begin{align}\label{eq23}
    \hat{U}_g^\dd \hat{b}_{gl}^{(0)} \hat{U}_g = \hat{b}_{gl}^{(0)\prime} &= \hat{b}_{gl}^{(0)} \cos\theta - i \hat{b}_{gr}^{(0)} e^{i\phi}\sin\theta \\ \label{eq25}
    \hat{U}_g^\dd \hat{b}_{gr}^{(0)} \hat{U}_g = \hat{b}_{gr}^{(0)\prime} &= \hat{b}_{gr}^{(0)} \cos\theta - i \hat{b}_{gl}^{(0)} e^{-i\phi}\sin\theta .
\end{align}
Here, $\cos\theta$ is the transmissivity of the mirror and $\phi$ is its phase. Thus, for $\theta = 0$, the action of the unitary reduces to an identity channel and the output operators are identical to the inputs. Now, using Eq.\ (\ref{unruh1}) and (\ref{unruh2}), the output Unruh operators can be obtained as follows
\begin{align} \label{eqn:unruh1}
    \hat{c}_{\omega i}^{(0)\prime} &= \hat{c}_{\omega i}^{(0)}  + g^\star(\omega)  \cosh r_\omega \big( \hat{U}_g^\dd \hat{b}_{gi}^{(0)} \hat{U}_g - \hat{b}_{gi}^{(0)} \big) \\ \label{eqn:unruh2}
    \hat{c}_{\omega i}^\text{(ex)$\prime$} &= \hat{c}_{\omega i}^\text{(ex)} - g(\omega)  \sinh r_\omega \big( \hat{U}_g^\dd \hat{b}_{gi}^{(0)\dd} \hat{U}_g - \hat{b}_{gi}^{(0)\dd } \big) .
\end{align}
The output diamond modes can be straightforwardly substituted into Eq.\ (\ref{eqn:unruh1}) and (\ref{eqn:unruh2}) to obtain the expanded form of the output Unruh operators. Using Eq.\ (\ref{eqn:unruh1}) and (\ref{eqn:unruh2}), we can construct output Minkowski operators, which are related to the output Unruh operators via
\begin{align}\label{eqn:Mink1}
    \hat{a}_{kl}^\prime &= \int \D \omega \big( A_{k\omega} \hat{c}_{\omega l}^{(0)\prime} + B_{k\omega} \hat{c}_{\omega l}^\text{(ex)$\prime$} \big) \\ \label{eqn:Mink2}
    \hat{a}_{kr}^\prime &= \int \D \omega \big( B_{k\omega}\hat{c}_{\omega r}^{(0)\prime} + A_{k\omega}\hat{c}_{\omega r}^\text{(ex)$\prime$}\big)
\end{align}
where $A_{k\omega},B_{k\omega}$ are Bogoliubov coefficients taking the form
\begin{align}\label{17}
    A_{k\omega} &= \frac{4\sqrt{\Omega\kappa}}{a} \sinh r_\omega e^{2i\kappa} M(1+i\Omega,2,-4i\kappa) \\ \label{18}
    B_{k\omega} &= \frac{4\sqrt{\Omega\kappa}}{a} \cosh r_\omega e^{2i\kappa} M(1-i\Omega,2,-4i\kappa),
\end{align}
as derived in Appendix A. We model the inertial observer as detecting wavepackets of Minkowski modes, defined by 
\begin{align}\label{eqn:wavepacket}
    \hat{a}_{fi}^\prime &= \int \D k \:f_i(k; k_0,\sigma,t_0,z_0) \hat{a}_{ki}^\prime 
\end{align}
where
\begin{align}
    f_l(k) &= \mathcal{N}_k \sqrt{k}\exp\left[ - \frac{(k-k_0)^2}{4\sigma^2}- ikV_0 \right]  \\
    f_r(k) &= \mathcal{N}_k \sqrt{k}\exp \left[ - \frac{(k-k_0)^2}{4\sigma^2} - i kU_0 \right] 
\end{align}
and $\mathcal{N}_k$ is a normalisation constant. As with the interacting diamond mode, the functions $f(k)$ have approximately Gaussian profiles in the narrow bandwidth, high-frequency regime ($k_0 \gg \sigma$). In this limit, $k_0$ is the centre-frequency, $\sigma$ the bandwidth and $V_0$ and $U_0$ are the centre positions of the left- and right-moving modes respectively. We note here that performing a detection prior to the interaction yields zero particles, since the incoming modes are Minkowski vacua, $\hat{a}_{fi}$. Only after the interaction, which is described by Eq.\ (\ref{eqn:Rindlerwavepacketeqn})-(\ref{eqn:unruh2}), is there a non-trivial transformation on the input diamond modes as they are evolved through the circuit.

\textit{Particle Production}\textemdash
In analogy with the uniformly accelerated mirror, we expect that the presence of the finite-lifetime mirror will induce particle production from the vacuum state. The motion of the accelerated mirror rapidly changes the boundary conditions of incoming field modes, altering the Hamiltonian of the system and generating particle excitations from the vacuum \cite{unruh1976notes,davies1977radiation,brown2015does}. Similarly, the rapid birth and then abrupt death of the finite-lifetime mirror within the causal diamond should induce similar vacuum excitations. To our knowledge, such an interaction has not been studied within the paradigm of causal diamonds and quantum fields. 

To determine the output state after this interaction, we substitute the output Unruh operators into Eq.\ (\ref{eqn:wavepacket}), from which one can derive the vacuum expectation value of the particle number in a wavepacket of output Minkowski modes, $\smash{\mathcal{N}_{i}(f) = \langle 0_M | a_{fi}^{\prime\dd} a_{fi}^{\prime} \ranglem}$. 
\begin{figure}[h]
    \centering
    \includegraphics[scale=0.35]{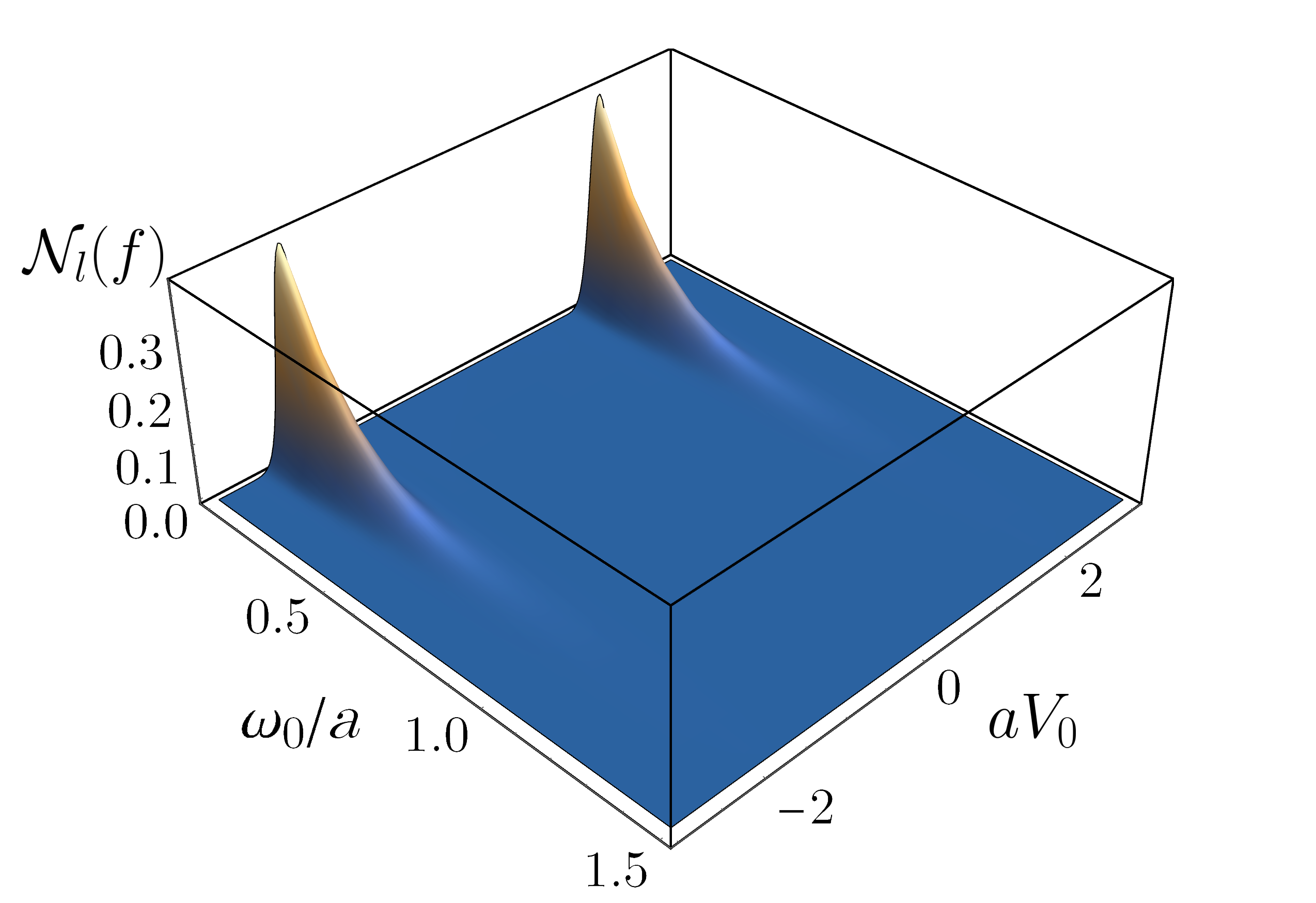}
    \caption{The particle number according to an inertial observer detecting the outgoing right-moving modes, as a function of $\omega_0/a$ and $aU_0$. The outgoing particle distributions are centred around $aU_0 = \pm 2$, along the initial and final half null-rays of the diamond. The parameters used here are $k_0/a= 12, \sigma/a = 3.2,  \delta/a = 0.2, \theta = \pi/2, \phi = 0$.}
    \label{fig:particles1}
\end{figure}
For illustration, the particle number according to a Minkowski detector of the left-moving modes is given by
\begin{align}\label{eqn:N}
    \mathcal{N}_l(f) 
    =2( 1 - \cos\theta) \Big[ |A_{fgl}|^2  \mathcal{I}_s + |B_{fgl}|^2  \mathcal{I}_c \Big]
\end{align}
where
\begin{align}\label{28}
    A_{fgl} &= \int\D \omega\:g^\star(\omega) \cosh r_\omega \int \D k \:f_l(k) A_{k\omega} \\ \label{29}
    B_{fgl} &= \int \D \omega\:g(\omega) \sinh r_\omega \int\D k \:f_l(k) B_{k\omega} 
\end{align}
and
\begin{align}
    \mathcal{I}_c &= \int\D \omega\:|g(\omega)|^2 \cosh^2r_\omega \\
     \mathcal{I}_s &= \int \D \omega\:|g(\omega)|^2 \sinh^2r_\omega.
\end{align} 
\noindent A symmetrical expression exists for the particle number of the outgoing right-moving modes. Furthermore, it should understood that within our (1+1)-dimensional model, the detector will register an identical particle count when placed anywhere along the half null-ray $U = \text{constant}$ after the interaction, irrespective of the $V$-coordinate. This intuition follows for a detector of the right-moving modes. The particle distribution of the outgoing left-moving modes is shown in Fig.\ \ref{fig:particles1}. We discover that the outgoing particles are concentrated at two peaks centred at the initial and final half null-rays of the diamond. This corresponds with our intuition about the rapid introduction and removal of the mirror, acting as a boundary for the modes. The spatial smearing in Fig.\ \ref{fig:particles1} is due to the imperfect spatial resolution of the detectors, which have a finite spread in frequency. The particle count decays with the centre-frequency of the reflected diamond mode in a similar manner to the Planckian thermal spectrum inside the diamond. Furthermore, the results are symmetric for the left- and right-moving output modes. 

\textit{Multi-partite entanglement}\textemdash
The unique distribution of particles produced by the finite-lifetime interaction motivates us to consider the entanglement structure of the output state. The finite-lifetime mirror passively mixes the incoming diamond modes via the unitary beamsplitter, Eq.\ (\ref{UG}), which does not create entanglement between them. Hence, discovering entanglement in the outgoing particle distributions would reveal the underlying entanglement in the vacuum state. The detection of vacuum entanglement (which has never been directly observed) would vindicate many theoretical predictions of relativistic QFT, and is closely related to phenomena such as the Unruh effect and Hawking radiation. Therefore, the finite-lifetime mirror may function as a stepping stone towards experimentally feasible proposals for observing this entanglement. 

To quantify the entanglement of the output state, we calculate the entanglement of formation (EoF) between a given pair of output Minkowski modes. Recent progress has been made in deriving analytical expressions for the EoF of two-mode Gaussian states, which has also been proven to be a more faithful measure of entanglement than say, the logarithmic negativity \cite{tserkis2017quantifying,tserkis2019quantifying}. By measuring the quadratures of the output modes, the inertial observer can construct the covariance matrix, 
\begin{align}
    \sigma_{ij} = \frac{1}{2}\langle \{ \hat{x}_i, \hat{x}_j\}\rangle 
\end{align}
which, for two-mode Gaussian states with zero mean value (here, we calculate the EoF for a given pair of output Minkowski modes) fully characterises the state. Here, $\hat{x}_i$ are proportional to the second-order moments of the quadrature field operators and $\{\:,\:\}$ is the anti-commutator \cite{weedbrook2012gaussian}. After calculating the covariance matrix $\sigma$, one obtains the EoF by the following formula \cite{bennett1996mixed,wolf2004gaussian,ivan2008entanglement,marian2008entanglement}, 
\begin{align}
    \mathcal{E}_F(\sigma) &= \inf_{\sigma_{p_i}} \left\{\mathcal{H} \left[\sigma_{p_i} (r) \right] |\sigma= \sigma_{p_i} + \phi_i \right\}
\end{align}
where $\sigma_p$ is the pure, two-mode squeezed vacuum characterised by the two-mode squeezing parameter $r$, $\phi\geq 0$ is a positive, semi-definite matrix, and $\mathcal{H}$ is the entropy of entanglement of $\sigma_p$ \cite{holevo1999capacity}
\begin{align}
    \mathcal{H}\left[\sigma_p(r) \right] = \cosh^2 r \log_2(\cosh^2 r) - \sinh^2 r\log_2(\sinh^2r) .
\end{align}
Computing the EoF reduces to an optimisation problem: finding $\sigma_p$ with the smallest $\mathcal{H}$ that can be transformed via local operations and classical communication into $\sigma$ \cite{bennett1996mixed,tserkis2019quantifying}. We utilise the \textsc{Mathematica} algorithm developed in \cite{tserkis2019quantifying} to evaluate the EoF between the output modes. 

We consider the detector configuration shown in Fig.\ \ref{fig:schematic}. Two inertial detectors are situated on either side of the mirror, detecting the outgoing Minkowski modes centred along and nearby the initial and final half null-rays of the diamond. If the particles detected along a given ray are entangled with those detected along the other three rays, this would signify the existence of genuine, multi-partite entanglement in the output state.  

\begin{figure*}[t]
  \includegraphics[width=1.0\linewidth]{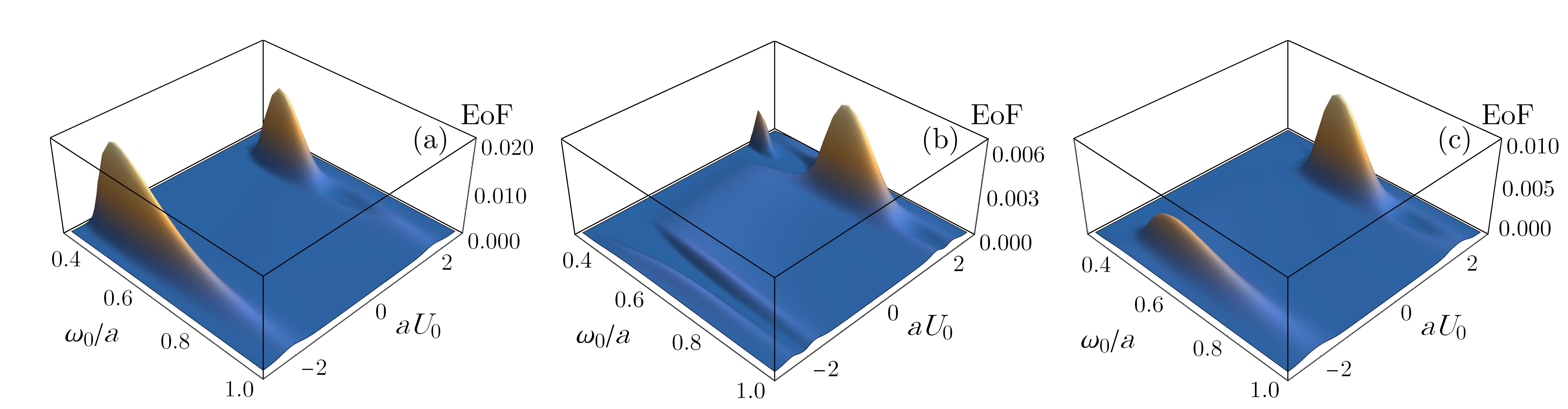}
  \caption{EoF between the detector of the left-moving modes fixed at the final half null-ray with the detector of the right-movers as it passes through the initial and final half null-rays. The regions of strongest entanglement are centred near the half null-rays, although for certain frequencies, the EoF bifurcates into two peaks before reviving once more. The EoF peaks also swap between the initial and final rays for different values of the phase. We have used the parameters $k_0/a = 8, \sigma/a  =3.2, \delta/a = 0.11, \theta = \pi/2$ and (a) $\phi = 0$, (b) $\phi = \pi/4$ and (c) $\phi = \pi/2$.}
  \label{fig:Combined1}
\end{figure*}

To demonstrate this, we study two cases. Fig.\ \ref{fig:Combined1} displays the spatial distribution of the EoF between the left-moving Minkowski modes detected at the final half null-ray and the right-moving modes, as a function of the centre frequency of the reflected mode, $\omega_0$ (that is, for narrow bandwidth modes $\omega_0\gg \delta$). The regions of entanglement are centred at the initial and final half null-rays of the diamond, where the outgoing right-moving particles are detected. We observe a complex entanglement structure between the outgoing modes, where (i) the region of strongest entanglement swaps between the final-final pair of modes and the final-initial pair for different values of $\phi$, and (ii) the entanglement exhibits frequency-dependent oscillations, and even a bifurcation and revival for certain pairs of modes, as $\omega_0/a$ is increased. This behaviour suggests that the correlations are distributed between different frequencies in a complicated manner. In general, the total amount of entanglement decays with $\omega_0/a$, since the quantum state inside the diamond begins to resemble the Minkowski vacuum, yielding fewer particle excitations \cite{su2019decoherence}. 

\begin{figure*}[t]
  \includegraphics[width=1.0\linewidth]{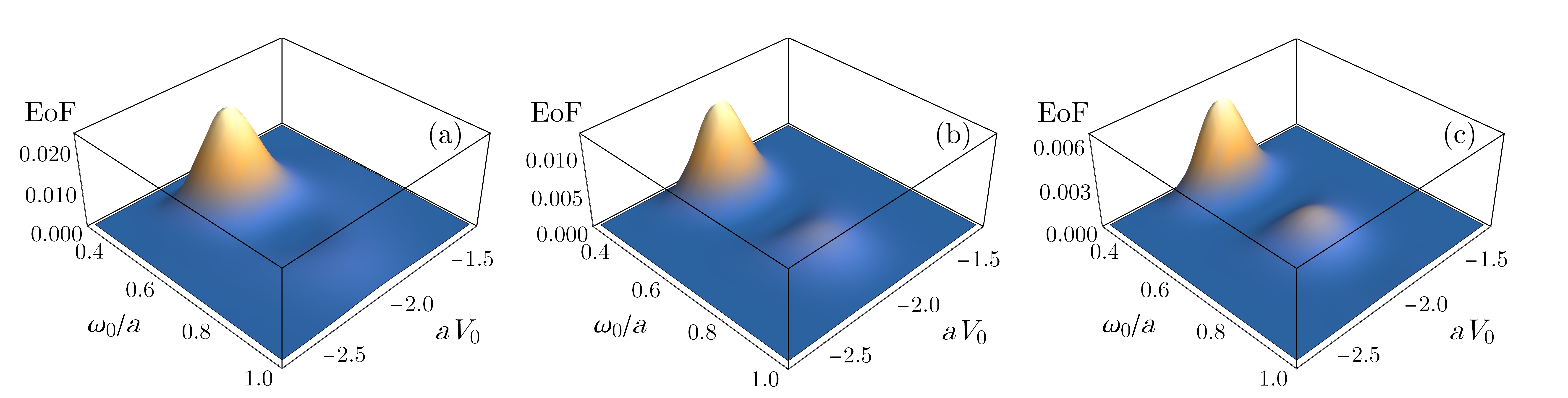}
  \caption{EoF between the detector of the left-moving modes at the final half null-ray with the another detector of the left-movers scanning through the initial half null-ray. The EoF decays in a non-trivial manner with increasing $\omega_0/a$. We have used the same parameters as above with (a) $k_0/a = 8$, (b) $k_0/a = 12$ and (c) $k_0/a = 16$. The EoF between modes moving in the same direction is $\phi$-independent.}
  \label{fig:Combined2}
\end{figure*}

Fig.\ \ref{fig:Combined2} shows the EoF distribution between the left-moving Minkowski modes detected at the final half null-ray while the other detector scans through the initial ray on the same side. In a similar manner to the left- and right-moving entanglement, the EoF peaks along the initial half null-ray, where the particles are predominantly concentrated. We also observe a decay and revival of the EoF, for similar reasons as discussed previously. Note that the detectors of modes moving in the same direction possess a small but non-vanishing commutator. We were careful to only consider cases where the effect on the particle number was negligible and verified that orthogonalising the two detectors using a Schmidt decomposition preserved strong entanglement \cite{onoe2018particle,onoe2019universal}. 
These calculations were repeated for a detector situated along all four of the diamond's null-rays, where we found entanglement with the modes centred along the other three rays. This verifies that the entanglement of the outgoing particles is genuinely multi-partite. Physical detection of this entanglement would be a signature of vacuum entanglement. Within the diamond reference frame, the unitary beamsplitter transformation, Eq.\ (\ref{UG}), does not entangle the incoming diamond modes, $\hat{b}_{gi}^{(0)}$ (i.e.\ describing thermal states). Hence, we infer that the entanglement of the outgoing particles originates from that already existing within the Minkowski vacuum state. Furthermore, we only considered cases where the overlap between Minkowski detectors of the modes moving in the same direction was negligible (that is, negating any spurious particle counts arising due to the Gaussian tails). 

\textit{Maximising bi-partite entanglement}\textemdash An important question is what parameter regimes maximise entanglement in the output state. Intuitively, we expect stronger entanglement between lower-frequency diamond modes (larger particle production) similarly detected by low-frequency Minkowski detectors. Furthermore, by utilising narrow bandwidth (in frequency) detectors, the individual particle distributions centred along the diamond's null-rays become spatio-temporally unresolvable, so that the observed entanglement is bi-partite between the left- and right-sides of the mirror. 
\begin{figure}[h]
    \centering
    \includegraphics[scale=0.335]{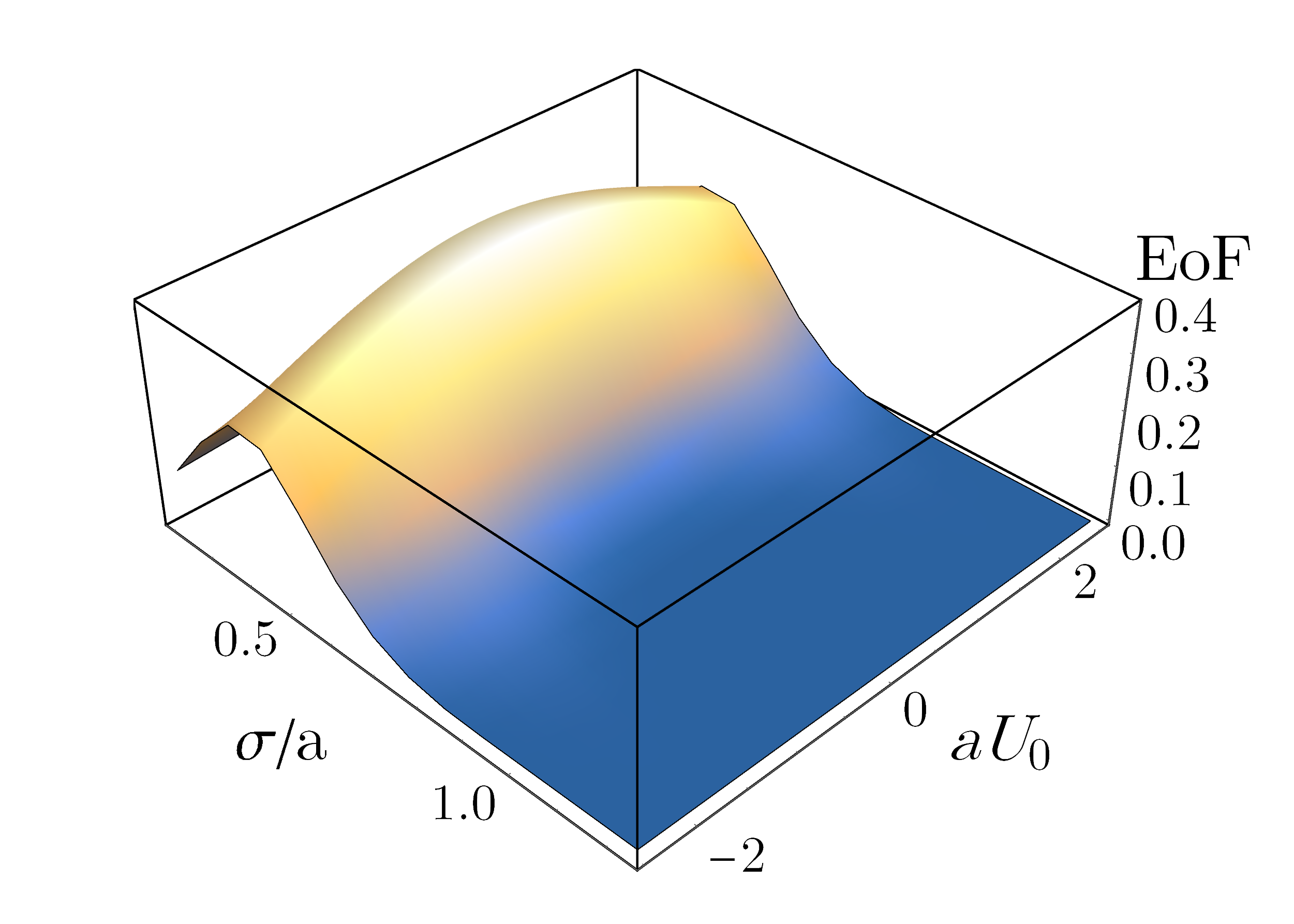}
    \caption{Entanglement of formation between the left- and right-moving Minkowski wavepacket modes, centred in-between the initial and final rays. For small $\sigma/a$, the entanglement peaks are indistinguishable, becoming increasingly so for larger $\sigma/a$. Here, we have used $k_0/a = 0.02, \omega_0/a = 0.01, \delta/a  = 0.4, \phi = 0, \theta = \pi/2, av_c = 0$.}
    \label{fig:3D2}
\end{figure}
Fig.\ \ref{fig:3D2} shows the spatial distribution of the EoF between the left- and right-moving Minkowski modes, for detectors situated in between the initial and final null-rays, as a function of the detector bandwidth $\sigma/a$. The maximum EoF is significantly larger than the multi-partite regime since the detectors count the individual particle distributions as a single unresolved peak. The optimal $\sigma/a$ is not arbitrarily small, indicating a trade-off between the spatial smearing of the wavepackets and the detection of the Minkowski modes with strongest entanglement.

One can also quantify the amount of entanglement by the variances of the quadrature correlations in the output (two-mode Gaussian) state, which is a more physically accessible measure. For a two-mode squeezed state $\sqrt{\mathcal{V}_X \mathcal{V}_P} < 1$
where $\mathcal{V}_X = \langle (X_0^\alpha - X_0^\beta)^2\rangle/2$ and $\mathcal{V}_P = \langle (P_0^\alpha + P_0^\beta)^2\rangle/2$. 
After applying symplectic transformations (phase rotations and local squeezing) to reduce the covariance matrix to the standard form, we find an optimal result of $\sqrt{\mathcal{V}_X \mathcal{V}_P} \sim 0.57$. This amount of quadrature squeezing is well within current, experimentally observable limits. One may consider this as an advantage of our approach over perturbative methods of entanglement harvesting, where the amount of detected entanglement is generally on the order of the weak detector-field coupling strength, $\mathcal{O}(\lambda^2)$ \cite{reznik2005violating,salton2015acceleration,ver2009entangling}. 

\textit{Finite energy fluxes}\textemdash
Finally, we consider the energy flux detected by an inertial observer after the finite-lifetime interaction between the diamond observer and incoming modes, and compare this with results found in the literature. It is well-known that the energy flux from an eternal, uniformly accelerating mirror is divergent \cite{frolov1979quantum}. An important question is whether the same divergences appear for the finite-lifetime mirror. One might expect that a mirror reflecting all frequencies of the input field would yield a divergent result, as recently shown using the circuit model for the uniformly accelerating mirror \cite{su2017quantum}. However in our model, the finite-lifetime mirror reflects a Gaussian wavepacket of diamond modes, which functionally switches the interaction on and off within the bounded causal diamond. 

\begin{figure}[h]
    \centering
    \includegraphics[scale=0.5]{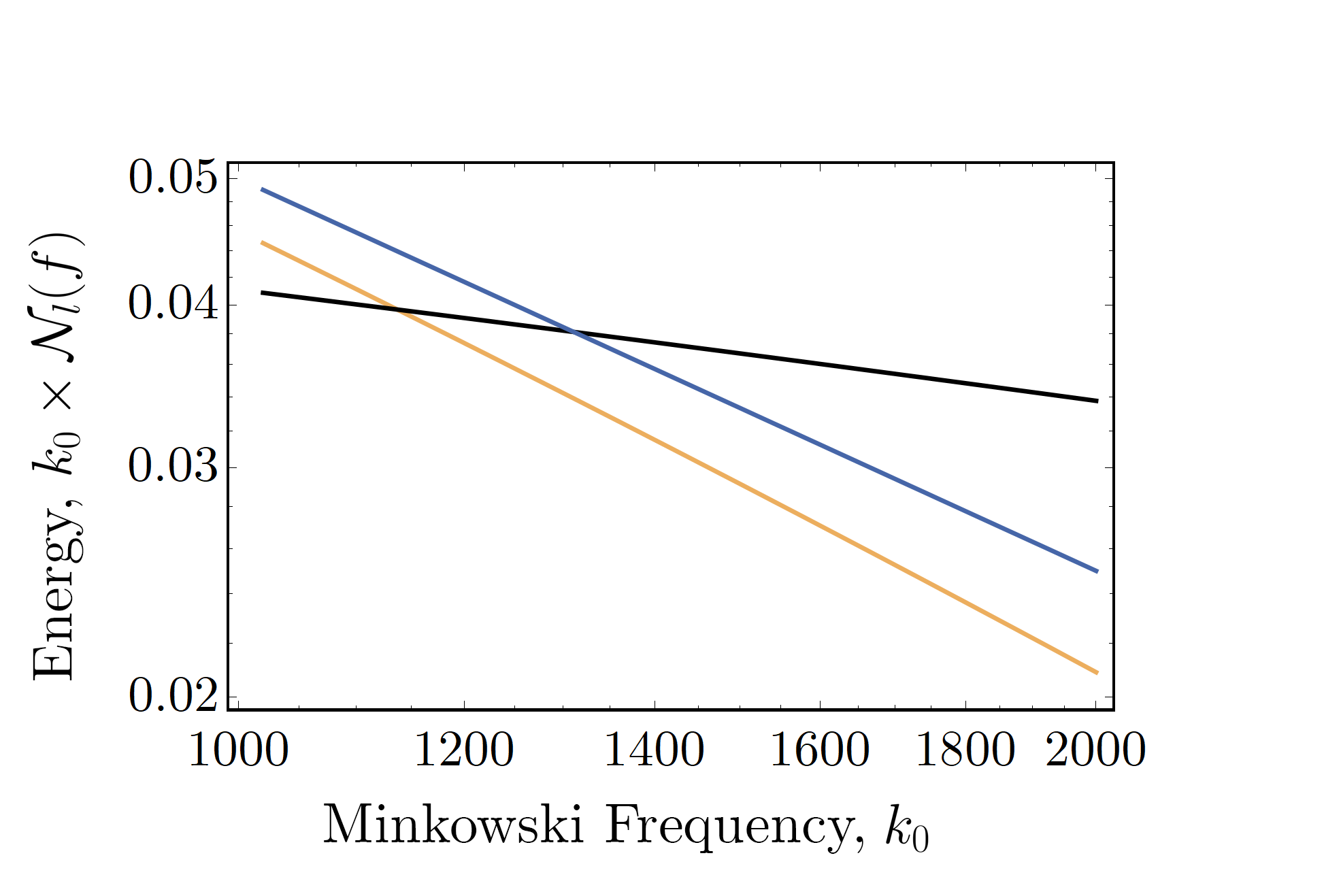}
    \caption{Log-log plot of the energy detected along the final half null-ray of the left-moving modes, as a function of the centre-frequency of the detected mode. The three curves represent (blue) a fitted function decaying as $1/k_0$, the energy of outgoing modes per $k_0$ with (black) $\delta/a = 0.1$, and (orange) $\delta/a = 0.2$. For an interaction with $\delta/a = 0.2$, the energy per mode decays faster than $1/k_0$, indicating the total energy flux is finite. The other parameters used are $\omega_0/a  = 5, aV_0 = 2,\theta = \pi/2, \sigma/a = 1$. }
    \label{fig:energy}
\end{figure}

To demonstrate the convergence of the total energy radiated by the mirror, it must be shown that the energy per wavepacket mode as detected by inertial observers decays faster than $1/k_0$ at high frequencies. This result would show that the total energy flux, integrated over all Minkowski modes, converges to a finite value. Now, in the high-frequency regime, the double integrals in Eq.\ (\ref{eqn:N}) become computationally expensive to calculate. An alternative approach is to decompose the Minkowski wavepacket operator as follows \cite{su2019decoherence,rohde2007spectral,onoe2018particle}, 
\begin{align}
     \hat{a}_{fl} &= \left( \hat{a}_{fl} - \big[\hat{a}_{fl}, \hat{b}_{gl}^{(0)\dd} \big] \hat{b}_{gl}^{(0)} - \big[ \hat{b}_{gl}^{(0)}, \hat{a}_{fl} \big] \hat{b}_{gl}^{(0)\dd} \right) \nonumber \\
     & + \left( \big[ \hat{a}_{fl}, \hat{b}_{gl}^{(0)\dd} \big] \hat{b}_{gl}^{(0)} + \big[ \hat{b}_{gl}^{(0)}, \hat{a}_{fl} \big] \hat{b}_{gl}^{(0)\dd} \right) 
\end{align}
where we have considered the left-moving modes without loss of generality. Since the first bracketed term is orthogonal to $\hat{a}_{fl}$ and $\hat{b}_{gl}$, only the second bracketed term is affected by the beamsplitter unitary, $\hat{U}_g$. The output Minkowski operator is thus:
\begin{align}
 \hat{a}_{fl}^{\dd\prime} &= \left( \hat{a}_{fl}^\dd - \big[ \hat{a}_{fl} , \hat{b}_{gl}^{(0)\dd} \big] \hat{b}_{gl}^{(0)} - \big[ \hat{b}_{gl}^{(0)} , \hat{a}_{fl} \big] \hat{b}_{gl}^{(0)\dd} \right) \nonumber \\
 & + \big[ \hat{a}_{fl}, \hat{b}_{gl}^{(0)\dd} \big] \left( \cos\theta \hat{b}_{gl}^{(0)} + \sin\theta \hat{b}_{gl}^{(0)} \right) \nonumber \\
 & + \big[ \hat{b}_{gl}^{(0)}, \hat{a}_{fl} \big] \left( \cos\theta \hat{b}_{gl}^{(0)\dd} + \sin\theta \hat{b}_{gl}^{(0)\dd} \right) .
\end{align}
It is straightforward to show that the particle number is given by
\begin{align}
    \mathcal{N}_l(f) &= 2(1 - \cos\theta) \big|\big[ \hat{a}_{fl}, \hat{b}_{gl}^{(0)\dd} \big] \big|^2 \big( \mathcal{I}_c + \mathcal{I}_s \big) .
\end{align}
This expression is equivalent to Eq.\ (\ref{eqn:N}). However rather than evaluating the double integrals in the commutator, we can make the association:
\begin{align}\label{44}
    \big[ \hat{a}_{fl}, \hat{b}_{gl}^{(0)\dd} \big] = \langle g_\mathfrak{G}^{(0)}(V), f_\mathfrak{G}(V) \rangle 
\end{align}
which is the Klein-Gordon inner product of the diamond wavepacket mode function and the Minkowski wavepacket mode function in position-space. Both $g_\mathfrak{G}^{(0)}(V)$ and $f_\mathfrak{G}(V)$ are obtained by convolving the respective single-frequency mode functions with a Gaussian wavepacket, expressed as
\begin{align}
     g_\mathfrak{G}^{(0)} (V) &=  \left( \frac{\delta^2}{2\pi\omega_0^2} \right)^{1/4} e^{ - V^{(0)} \left (V^{(0)} \delta^2 + i \omega_0 \right)}\\
    f_\mathfrak{G}(V) &=  \left( \frac{\sigma^2}{2\pi k_0^2} \right)^{1/4} e^{ - (V-V_0) \left( (V-V_0) \sigma^2 + i k_0 \right) } 
\end{align}
where
\begin{align}
    V^{(0)} &= a^{-1} \ln \left( \frac{1 + aV/2}{1 - a V/2} \right) .
\end{align}
After calculating $\langle g_\mathfrak{G}^{(0)}(V), f_\mathfrak{G}(V)\rangle$, we plot the energy flux $(k_0 \times \mathcal{N}_l)$ as a function of the centre-frequency of the detected Minkowski wavepacket mode, shown in Fig.\ \ref{fig:energy}. We find that for sufficiently large $\delta/a$, the energy radiated by the mirror decays faster than $1/k_0$.  
In this regime, integrating over all modes of the field (and taking into account the low-frequency energy convergence, below) yields a finite result. The large $\delta/a$ regime amounts to an interaction which is sufficiently localised within the diamond, due to the Fourier-transform relationship between frequency and time. This result is corroborated by previous works which obtained finite energy fluxes for an interaction smoothly switched on and off \cite{obadia2001notes,obadia2003uniformly}. For smaller $\delta/a$, the energy per mode decays slower than $1/k_0$. The numerical methods we utilised could not probe significantly higher frequencies, and hence the energy convergence in this regime is inconclusive. 

In the low Minkowski-frequency limit ($|z|\ll 1)$, the energy also converges. To show this, we utilise the asymptotic form of $M(a,b,z) \to 1$ for $|z|\ll 1$ \cite{abramowitz1999ia}, such that the Bogoliubov coefficients Eq.\ (\ref{28}) and (\ref{29}) reduce to
\begin{align}\label{43}
    A_{k\omega} &\simeq \frac{4\sqrt{\Omega\kappa}}{a} \sinh r_\omega e^{2i\kappa} \\ \label{44}
    B_{k\omega} &\simeq \frac{4\sqrt{\Omega\kappa}}{a} \cosh r_\omega e^{2i\kappa}.
\end{align}
From Eq.\ (\ref{43}) and (\ref{44}) and in the limit $|k|\ll 1$, we notice that the Bogoliubov coefficients Eq.\ (\ref{28}) and (\ref{29}), and hence the particle number, vanish, implying that $\mathcal{E}(f) \to 0$ as $k_0\to 0$. Thus, the problem of energy divergences is resolved by modelling the mirror as reflecting a Gaussian wavepacket of diamond modes.

\textit{Conclusion}\textemdash In this paper, we have studied the interaction between a finite-lifetime mirror confined to a bounded causal diamond and incoming field modes. Inertial observers in the Minkowski vacuum observe particle production along the half null-rays of the diamond's lifetime, which possess genuine, multi-partite entanglement with a highly non-trivial and novel spatio-temporal and frequency dependence. We inferred that this entanglement originated from that pre-existing in the vacuum state of the quantum field, since the mirror does not entangle the incoming modes in the diamond reference frame. 

The field quantisation presented here offers a natural approach to localising field modes and general unitary interactions with quantum objects (such as those commonly used in quantum optics), without requiring compactly supported switching functions or boundary conditions. As stated previously, strictly localised modes have foreseeable applications in problems where causal order and localisation of the modes is important. For example, interactions inside multiple diamond regions could be used to simulate a quantum switch, that is, an interaction that occurs in a coherent superposition of temporal orders \cite{brukner2014quantum,chiribella2013quantum,araujo2014computational}. Other probes of the causal structure of QFT, such as the violation of causal inequalities \cite{branciard2015simplest}, may be accessible by constructing quantum communication protocols between spatio-temporally localised diamond observers \cite{ho2018violation}. As already mentioned, the diamond region is conformal to the static patch of de Sitter spacetime \cite{tian2005sitter}. An interesting direction would be to rigorously derive the coordinate transformations between these two spacetimes.

\textit{Acknowledgments}\textemdash 
The authors would like to acknowledge support from the Australian
Research Council Centre of Excellence for Quantum
Computation and Communication Technology (Project
No. CE170100012) and and DECRA Grant DE180101443.

\appendix
\begin{widetext}
\section{Derivation of $A_{k\omega}$ and $B_{k\omega}$}
\noindent Here, we outline the derivation of the Bogoliubov coefficients, Eq.\ (\ref{17}) and (\ref{18}). The left-moving Unruh mode functions can be expressed in terms of the diamond mode functions, 
\begin{align}\label{A1}
     G_{\omega}^{(0)}(V) &= \cosh r_\omega g_{\omega}^{(0)}(V)  + \sinh r_\omega g_{\omega }^\text{(ex)$\star$}(V) \\ \label{A2} 
    {G}^\text{(ex)}_{\omega }(V) &= \cosh r_\omega g_{\omega }^\text{(ex)}(V) + \sinh r_\omega  g_{\omega}^{(0)\star} (V)
\end{align}
Using the relation between the diamond modes and the Minkowski modes, 
\begin{align} 
    g_{\omega }^{(0)} (V) &= \int_0^\infty \D k \left( \alpha_{\omega k}^{(0)} u_{k}(V) + \beta_{\omega k}^{(0)} u_k^\star(V) \right) \\ \label{61}
    g_\omega^\text{(ex)}(V) &= \int_0^\infty \D k \left( \alpha_{\omega k}^\text{(ex)} u_k(V) + \beta_{\omega k}^\text{(ex)} u_k^\star( V) \right) 
\end{align}
and substituting them into Eq.\ (\ref{A1}) and (\ref{A2}), we find that,
\begin{align} \label{eqn:a5}
G_\omega^{(0)}(V) &= \frac{1-e^{-2\pi\Omega}}{\sqrt{1-e^{-2\pi\Omega}}} \int_0^\infty \D k \:\alpha_{\omega k}^{(0)} u_k \\ \label{eqn:a6}
{G}^\text{(ex)}_\omega(V) &= \frac{-e^{-\pi\Omega}(1-e^{-2\pi\Omega})}{\sqrt{1-e^{-2\pi\Omega}}} \int_0^\infty \D k \:\beta_{\omega k}^{(0)\star} u_k 
\end{align}
where by construction, $    \beta_{\omega k}^{(0)} = - e^{-\pi\Omega} \alpha_{\omega k}^\text{(ex)$\star$} 
$ and $\beta_{\omega k}^\text{(ex)} = -e^{-\pi\Omega}\alpha_{\omega k}^{(0)\star}$. Recalling that the 
Unruh and Minkowski modes share the same vacuum, then by definition, \begin{align}
    G^{(0)}_\omega(V) &= \int_0^\infty \D k \:A_{k\omega}u_k(V) \\
    G^\text{(ex)}_\omega(V) &=\int_0^\infty \D k \:B_{k\omega}u_k(V) 
\end{align}
so substituting Eq.\ (\ref{alpha}) and (\ref{beta}) into Eq.\  (\ref{eqn:a5}) and (\ref{eqn:a6}) and equating terms, we obtain
\begin{align}
    A_{k\omega} &= \frac{4\sqrt{\Omega\kappa}}{a} \sinh r_\omega e^{2i\kappa}M(1+i\Omega,2,-4i\kappa) \\
    B_{k\omega} &= \frac{4\sqrt{\Omega\kappa}}{a} \cosh r_\omega e^{2i\kappa} M(1-i\Omega,2,-4i\kappa) .
\end{align}

\section{Covariance Matrix Elements}
\subsection{Left-Right Entanglement}
\noindent Let us define the vector operator 
\begin{align}
    \hat{\textbf{x}} &= (\hat{X}_l(0), \hat{X}_l(\pi/2), \hat{X}_r(0) ,\hat{X}_r(\pi/2) )^T 
\end{align}
where $\hat{X} (0) = a + a^\dd$, $\hat{X}(\pi/2) = -i(a - a^\dd)$ are the quadrature operators commonly associated with position and momentum. The two-mode covariance matrix has elements given by 
\begin{align}
    \sigma_{ij} &= \frac{1}{2} \langlem  \left\{ \delta \hat{x}_i ,\delta \hat{x}_j \right\}  \ranglem 
\end{align}
where $\delta \hat{x}_i = \hat{x}_i - \langle \hat{x}_i \rangle$. When the expectation values of the quadrature amplitudes are zero, which is true of the system we are considering, $\hat{\sigma}$ contains full information about the state. For left-right entanglement, the non-zero terms which contribute to the covariance matrix are given by
\begin{align}
\langlem \hat{a}_{fl}{\dd} \hat{a}_{f'l}\ranglem  &= 2(1 - \cos\theta) \Big[ |A_{fgl}|^2 \mathcal{I}_s + |B_{fgl} |^2\mathcal{I}_s \Big] \nonumber \vphantom{\Big]}\\
       \langlem \hat{a}_{fl} \hat{a}_{f'l} \ranglem &= -2 ( 1 - \cos \theta) A_{fgl} B_{f'g'l} \nonumber \vphantom{\Big]}\\
       \langlem  \hat{a}_{fr}^{\dd} \hat{a}_{f'r} \ranglem &= 2(1 - \cos\theta) \Big[ |A_{fgr}|^2 \mathcal{I}_s + |B_{fgr}|^2\mathcal{I}_s \Big] \nonumber \vphantom{\Big]}\\
        \langlem \hat{a}_{fr} \hat{a}_{f'r} \ranglem &= -2 ( 1- \cos \theta) A_{fgr} B_{f'g'r} \nonumber \vphantom{\Big]}\\
        \langlem \hat{a}_{fl} \hat{a}_{f'r} \ranglem &= -i \sin\theta \Big[ e^{i\phi} A_{fgl}A_{f'g'r} + e^{-i\phi}B_{fgl}B_{f'g'r} \Big] .
\end{align}

\subsection{Left-Left Entanglement} 
\noindent The only difference between analysis of the left-left and left-right entanglement are the non-zero terms,
\begin{align}
    \langlem \hat{a}^+_{fl} \hat{a}^-_{f'l} \ranglem  &= - ( 1 - \cos \theta) \Big[ A_{fgl}^+ B_{f'g'l}^- + B_{fgl}^+ A_{f'g'l}^- \Big] \nonumber \vphantom{\Big]} \\
   \langlem  \hat{a}^{+\dd}_{fl} \hat{a}^-_{f'l} \ranglem &= 2 ( 1 - \cos \theta) \Big[ A_{fgl}^{+\star} A_{f'g'l}^- + B_{fgl}^{+\star} B_{f'g'l}^- \Big] \nonumber \vphantom{\Big]}
\end{align}
where the superscripts $`\pm'$ denote detectors situated along the final and initial null-rays respectively. Analogous expressions follow for entanglement between right-moving modes at the final and initial rays. 
\\\\

\end{widetext}
\bibliography{References.bib}

\begin{thebibliography}{51}%
\makeatletter
\providecommand \@ifxundefined [1]{%
 \@ifx{#1\undefined}
}%
\providecommand \@ifnum [1]{%
 \ifnum #1\expandafter \@firstoftwo
 \else \expandafter \@secondoftwo
 \fi
}%
\providecommand \@ifx [1]{%
 \ifx #1\expandafter \@firstoftwo
 \else \expandafter \@secondoftwo
 \fi
}%
\providecommand \natexlab [1]{#1}%
\providecommand \enquote  [1]{``#1''}%
\providecommand \bibnamefont  [1]{#1}%
\providecommand \bibfnamefont [1]{#1}%
\providecommand \citenamefont [1]{#1}%
\providecommand \href@noop [0]{\@secondoftwo}%
\providecommand \href [0]{\begingroup \@sanitize@url \@href}%
\providecommand \@href[1]{\@@startlink{#1}\@@href}%
\providecommand \@@href[1]{\endgroup#1\@@endlink}%
\providecommand \@sanitize@url [0]{\catcode `\\12\catcode `\$12\catcode
  `\&12\catcode `\#12\catcode `\^12\catcode `\_12\catcode `\%12\relax}%
\providecommand \@@startlink[1]{}%
\providecommand \@@endlink[0]{}%
\providecommand \url  [0]{\begingroup\@sanitize@url \@url }%
\providecommand \@url [1]{\endgroup\@href {#1}{\urlprefix }}%
\providecommand \urlprefix  [0]{URL }%
\providecommand \Eprint [0]{\href }%
\providecommand \doibase [0]{http://dx.doi.org/}%
\providecommand \selectlanguage [0]{\@gobble}%
\providecommand \bibinfo  [0]{\@secondoftwo}%
\providecommand \bibfield  [0]{\@secondoftwo}%
\providecommand \translation [1]{[#1]}%
\providecommand \BibitemOpen [0]{}%
\providecommand \bibitemStop [0]{}%
\providecommand \bibitemNoStop [0]{.\EOS\space}%
\providecommand \EOS [0]{\spacefactor3000\relax}%
\providecommand \BibitemShut  [1]{\csname bibitem#1\endcsname}%
\let\auto@bib@innerbib\@empty
\bibitem [{\citenamefont {Connes}\ and\ \citenamefont
  {Rovelli}(1994)}]{connes1994neumann}%
  \BibitemOpen
  \bibfield  {author} {\bibinfo {author} {\bibfnamefont {A.}~\bibnamefont
  {Connes}}\ and\ \bibinfo {author} {\bibfnamefont {C.}~\bibnamefont
  {Rovelli}},\ }\href@noop {} {\bibfield  {journal} {\bibinfo  {journal}
  {Classical and Quantum Gravity}\ }\textbf {\bibinfo {volume} {11}},\ \bibinfo
  {pages} {2899} (\bibinfo {year} {1994})}\BibitemShut {NoStop}%
\bibitem [{\citenamefont {Martinetti}\ and\ \citenamefont
  {Rovelli}(2003)}]{martinetti2003diamond}%
  \BibitemOpen
  \bibfield  {author} {\bibinfo {author} {\bibfnamefont {P.}~\bibnamefont
  {Martinetti}}\ and\ \bibinfo {author} {\bibfnamefont {C.}~\bibnamefont
  {Rovelli}},\ }\href@noop {} {\bibfield  {journal} {\bibinfo  {journal}
  {Classical and Quantum Gravity}\ }\textbf {\bibinfo {volume} {20}},\ \bibinfo
  {pages} {4919} (\bibinfo {year} {2003})}\BibitemShut {NoStop}%
\bibitem [{\citenamefont {Birrell}\ and\ \citenamefont
  {Davies}(1984)}]{birrelldavies}%
  \BibitemOpen
  \bibfield  {author} {\bibinfo {author} {\bibfnamefont {N.~D.}\ \bibnamefont
  {Birrell}}\ and\ \bibinfo {author} {\bibfnamefont {P.}~\bibnamefont
  {Davies}},\ }\href@noop {} {\emph {\bibinfo {title} {Quantum fields in curved
  space}}},\ \bibinfo {series} {1}\ No.~\bibinfo {number} {7}\ (\bibinfo
  {publisher} {Cambridge university press},\ \bibinfo {year}
  {1984})\BibitemShut {NoStop}%
\bibitem [{\citenamefont {Unruh}(1976)}]{unruh1976notes}%
  \BibitemOpen
  \bibfield  {author} {\bibinfo {author} {\bibfnamefont {W.~G.}\ \bibnamefont
  {Unruh}},\ }\href@noop {} {\bibfield  {journal} {\bibinfo  {journal}
  {Physical Review D}\ }\textbf {\bibinfo {volume} {14}},\ \bibinfo {pages}
  {870} (\bibinfo {year} {1976})}\BibitemShut {NoStop}%
\bibitem [{\citenamefont {Crispino}\ \emph {et~al.}(2008)\citenamefont
  {Crispino}, \citenamefont {Higuchi},\ and\ \citenamefont
  {Matsas}}]{crispino2008unruh}%
  \BibitemOpen
  \bibfield  {author} {\bibinfo {author} {\bibfnamefont {L.~C.}\ \bibnamefont
  {Crispino}}, \bibinfo {author} {\bibfnamefont {A.}~\bibnamefont {Higuchi}}, \
  and\ \bibinfo {author} {\bibfnamefont {G.~E.}\ \bibnamefont {Matsas}},\
  }\href@noop {} {\bibfield  {journal} {\bibinfo  {journal} {Reviews of Modern
  Physics}\ }\textbf {\bibinfo {volume} {80}},\ \bibinfo {pages} {787}
  (\bibinfo {year} {2008})}\BibitemShut {NoStop}%
\bibitem [{\citenamefont {Alsing}\ and\ \citenamefont
  {Milonni}(2004)}]{alsing2004simplified}%
  \BibitemOpen
  \bibfield  {author} {\bibinfo {author} {\bibfnamefont {P.~M.}\ \bibnamefont
  {Alsing}}\ and\ \bibinfo {author} {\bibfnamefont {P.~W.}\ \bibnamefont
  {Milonni}},\ }\href@noop {} {\bibfield  {journal} {\bibinfo  {journal}
  {American Journal of Physics}\ }\textbf {\bibinfo {volume} {72}},\ \bibinfo
  {pages} {1524} (\bibinfo {year} {2004})}\BibitemShut {NoStop}%
\bibitem [{\citenamefont {Su}\ and\ \citenamefont
  {Ralph}(2016)}]{su2016spacetime}%
  \BibitemOpen
  \bibfield  {author} {\bibinfo {author} {\bibfnamefont {D.}~\bibnamefont
  {Su}}\ and\ \bibinfo {author} {\bibfnamefont {T.}~\bibnamefont {Ralph}},\
  }\href@noop {} {\bibfield  {journal} {\bibinfo  {journal} {Physical Review
  D}\ }\textbf {\bibinfo {volume} {93}},\ \bibinfo {pages} {044023} (\bibinfo
  {year} {2016})}\BibitemShut {NoStop}%
\bibitem [{\citenamefont {Schlieder}(1965)}]{schlieder1965some}%
  \BibitemOpen
  \bibfield  {author} {\bibinfo {author} {\bibfnamefont {S.}~\bibnamefont
  {Schlieder}},\ }\href@noop {} {\bibfield  {journal} {\bibinfo  {journal}
  {Communications in Mathematical Physics}\ }\textbf {\bibinfo {volume} {1}},\
  \bibinfo {pages} {265} (\bibinfo {year} {1965})}\BibitemShut {NoStop}%
\bibitem [{\citenamefont {Haag}\ and\ \citenamefont
  {Swieca}(1965)}]{haag1965does}%
  \BibitemOpen
  \bibfield  {author} {\bibinfo {author} {\bibfnamefont {R.}~\bibnamefont
  {Haag}}\ and\ \bibinfo {author} {\bibfnamefont {J.~A.}\ \bibnamefont
  {Swieca}},\ }\href@noop {} {\bibfield  {journal} {\bibinfo  {journal}
  {Communications in Mathematical Physics}\ }\textbf {\bibinfo {volume} {1}},\
  \bibinfo {pages} {308} (\bibinfo {year} {1965})}\BibitemShut {NoStop}%
\bibitem [{\citenamefont {Knight}(1961)}]{knight1961strict}%
  \BibitemOpen
  \bibfield  {author} {\bibinfo {author} {\bibfnamefont {J.~M.}\ \bibnamefont
  {Knight}},\ }\href@noop {} {\bibfield  {journal} {\bibinfo  {journal}
  {Journal of Mathematical Physics}\ }\textbf {\bibinfo {volume} {2}},\
  \bibinfo {pages} {459} (\bibinfo {year} {1961})}\BibitemShut {NoStop}%
\bibitem [{\citenamefont {Satz}(2007)}]{satz2007then}%
  \BibitemOpen
  \bibfield  {author} {\bibinfo {author} {\bibfnamefont {A.}~\bibnamefont
  {Satz}},\ }\href@noop {} {\bibfield  {journal} {\bibinfo  {journal}
  {Classical and Quantum Gravity}\ }\textbf {\bibinfo {volume} {24}},\ \bibinfo
  {pages} {1719} (\bibinfo {year} {2007})}\BibitemShut {NoStop}%
\bibitem [{\citenamefont {Louko}\ and\ \citenamefont
  {Satz}(2006)}]{louko2006often}%
  \BibitemOpen
  \bibfield  {author} {\bibinfo {author} {\bibfnamefont {J.}~\bibnamefont
  {Louko}}\ and\ \bibinfo {author} {\bibfnamefont {A.}~\bibnamefont {Satz}},\
  }\href@noop {} {\bibfield  {journal} {\bibinfo  {journal} {Classical and
  Quantum Gravity}\ }\textbf {\bibinfo {volume} {23}},\ \bibinfo {pages} {6321}
  (\bibinfo {year} {2006})}\BibitemShut {NoStop}%
\bibitem [{\citenamefont {Friis}\ \emph {et~al.}(2012)\citenamefont {Friis},
  \citenamefont {Huber}, \citenamefont {Fuentes},\ and\ \citenamefont
  {Bruschi}}]{friis2012quantum}%
  \BibitemOpen
  \bibfield  {author} {\bibinfo {author} {\bibfnamefont {N.}~\bibnamefont
  {Friis}}, \bibinfo {author} {\bibfnamefont {M.}~\bibnamefont {Huber}},
  \bibinfo {author} {\bibfnamefont {I.}~\bibnamefont {Fuentes}}, \ and\
  \bibinfo {author} {\bibfnamefont {D.~E.}\ \bibnamefont {Bruschi}},\
  }\href@noop {} {\bibfield  {journal} {\bibinfo  {journal} {Physical Review
  D}\ }\textbf {\bibinfo {volume} {86}},\ \bibinfo {pages} {105003} (\bibinfo
  {year} {2012})}\BibitemShut {NoStop}%
\bibitem [{\citenamefont {Bruschi}\ \emph {et~al.}(2012)\citenamefont
  {Bruschi}, \citenamefont {Fuentes},\ and\ \citenamefont
  {Louko}}]{bruschi2012voyage}%
  \BibitemOpen
  \bibfield  {author} {\bibinfo {author} {\bibfnamefont {D.~E.}\ \bibnamefont
  {Bruschi}}, \bibinfo {author} {\bibfnamefont {I.}~\bibnamefont {Fuentes}}, \
  and\ \bibinfo {author} {\bibfnamefont {J.}~\bibnamefont {Louko}},\
  }\href@noop {} {\bibfield  {journal} {\bibinfo  {journal} {Physical Review
  D}\ }\textbf {\bibinfo {volume} {85}},\ \bibinfo {pages} {061701} (\bibinfo
  {year} {2012})}\BibitemShut {NoStop}%
\bibitem [{\citenamefont {Brown}\ \emph {et~al.}(2015)\citenamefont {Brown},
  \citenamefont {del Rey}, \citenamefont {Westman}, \citenamefont {Le{\'o}n},\
  and\ \citenamefont {Dragan}}]{brown2015does}%
  \BibitemOpen
  \bibfield  {author} {\bibinfo {author} {\bibfnamefont {E.~G.}\ \bibnamefont
  {Brown}}, \bibinfo {author} {\bibfnamefont {M.}~\bibnamefont {del Rey}},
  \bibinfo {author} {\bibfnamefont {H.}~\bibnamefont {Westman}}, \bibinfo
  {author} {\bibfnamefont {J.}~\bibnamefont {Le{\'o}n}}, \ and\ \bibinfo
  {author} {\bibfnamefont {A.}~\bibnamefont {Dragan}},\ }\href@noop {}
  {\bibfield  {journal} {\bibinfo  {journal} {Physical Review D}\ }\textbf
  {\bibinfo {volume} {91}},\ \bibinfo {pages} {016005} (\bibinfo {year}
  {2015})}\BibitemShut {NoStop}%
\bibitem [{\citenamefont {Brown}\ and\ \citenamefont
  {Louko}(2015)}]{brown2015smooth}%
  \BibitemOpen
  \bibfield  {author} {\bibinfo {author} {\bibfnamefont {E.~G.}\ \bibnamefont
  {Brown}}\ and\ \bibinfo {author} {\bibfnamefont {J.}~\bibnamefont {Louko}},\
  }\href@noop {} {\bibfield  {journal} {\bibinfo  {journal} {Journal of High
  Energy Physics}\ }\textbf {\bibinfo {volume} {2015}},\ \bibinfo {pages} {61}
  (\bibinfo {year} {2015})}\BibitemShut {NoStop}%
\bibitem [{\citenamefont {Oreshkov}\ \emph {et~al.}(2012)\citenamefont
  {Oreshkov}, \citenamefont {Costa},\ and\ \citenamefont
  {Brukner}}]{oreshkov2012quantum}%
  \BibitemOpen
  \bibfield  {author} {\bibinfo {author} {\bibfnamefont {O.}~\bibnamefont
  {Oreshkov}}, \bibinfo {author} {\bibfnamefont {F.}~\bibnamefont {Costa}}, \
  and\ \bibinfo {author} {\bibfnamefont {{\v{C}}.}~\bibnamefont {Brukner}},\
  }\href@noop {} {\bibfield  {journal} {\bibinfo  {journal} {Nature
  communications}\ }\textbf {\bibinfo {volume} {3}},\ \bibinfo {pages} {1}
  (\bibinfo {year} {2012})}\BibitemShut {NoStop}%
\bibitem [{\citenamefont {Fulling}\ and\ \citenamefont
  {Davies}(1976)}]{fulling1976radiation}%
  \BibitemOpen
  \bibfield  {author} {\bibinfo {author} {\bibfnamefont {S.~A.}\ \bibnamefont
  {Fulling}}\ and\ \bibinfo {author} {\bibfnamefont {P.~C.}\ \bibnamefont
  {Davies}},\ }\href@noop {} {\bibfield  {journal} {\bibinfo  {journal}
  {Proceedings of the Royal Society of London. A. Mathematical and Physical
  Sciences}\ }\textbf {\bibinfo {volume} {348}},\ \bibinfo {pages} {393}
  (\bibinfo {year} {1976})}\BibitemShut {NoStop}%
\bibitem [{\citenamefont {Walker}(1985)}]{walker1985particle}%
  \BibitemOpen
  \bibfield  {author} {\bibinfo {author} {\bibfnamefont {W.}~\bibnamefont
  {Walker}},\ }\href@noop {} {\bibfield  {journal} {\bibinfo  {journal}
  {Physical Review D}\ }\textbf {\bibinfo {volume} {31}},\ \bibinfo {pages}
  {767} (\bibinfo {year} {1985})}\BibitemShut {NoStop}%
\bibitem [{\citenamefont {Carlitz}\ and\ \citenamefont
  {Willey}(1987)}]{carlitz1987reflections}%
  \BibitemOpen
  \bibfield  {author} {\bibinfo {author} {\bibfnamefont {R.~D.}\ \bibnamefont
  {Carlitz}}\ and\ \bibinfo {author} {\bibfnamefont {R.~S.}\ \bibnamefont
  {Willey}},\ }\href@noop {} {\bibfield  {journal} {\bibinfo  {journal}
  {Physical Review D}\ }\textbf {\bibinfo {volume} {36}},\ \bibinfo {pages}
  {2327} (\bibinfo {year} {1987})}\BibitemShut {NoStop}%
\bibitem [{\citenamefont {Su}\ \emph {et~al.}(2017)\citenamefont {Su},
  \citenamefont {Ho}, \citenamefont {Mann},\ and\ \citenamefont
  {Ralph}}]{su2017quantum}%
  \BibitemOpen
  \bibfield  {author} {\bibinfo {author} {\bibfnamefont {D.}~\bibnamefont
  {Su}}, \bibinfo {author} {\bibfnamefont {C.~M.}\ \bibnamefont {Ho}}, \bibinfo
  {author} {\bibfnamefont {R.~B.}\ \bibnamefont {Mann}}, \ and\ \bibinfo
  {author} {\bibfnamefont {T.~C.}\ \bibnamefont {Ralph}},\ }\href@noop {}
  {\bibfield  {journal} {\bibinfo  {journal} {New Journal of Physics}\ }\textbf
  {\bibinfo {volume} {19}},\ \bibinfo {pages} {063017} (\bibinfo {year}
  {2017})}\BibitemShut {NoStop}%
\bibitem [{\citenamefont {Su}\ and\ \citenamefont
  {Ralph}(2019)}]{su2019decoherence}%
  \BibitemOpen
  \bibfield  {author} {\bibinfo {author} {\bibfnamefont {D.}~\bibnamefont
  {Su}}\ and\ \bibinfo {author} {\bibfnamefont {T.~C.}\ \bibnamefont {Ralph}},\
  }\href@noop {} {\bibfield  {journal} {\bibinfo  {journal} {Physical Review
  X}\ }\textbf {\bibinfo {volume} {9}},\ \bibinfo {pages} {011007} (\bibinfo
  {year} {2019})}\BibitemShut {NoStop}%
\bibitem [{\citenamefont {Coleman}(1973)}]{coleman1973there}%
  \BibitemOpen
  \bibfield  {author} {\bibinfo {author} {\bibfnamefont {S.}~\bibnamefont
  {Coleman}},\ }\href@noop {} {\bibfield  {journal} {\bibinfo  {journal}
  {Communications in Mathematical Physics}\ }\textbf {\bibinfo {volume} {31}},\
  \bibinfo {pages} {259} (\bibinfo {year} {1973})}\BibitemShut {NoStop}%
\bibitem [{\citenamefont {De~Lorenzo}\ and\ \citenamefont
  {Perez}(2018)}]{de2018light}%
  \BibitemOpen
  \bibfield  {author} {\bibinfo {author} {\bibfnamefont {T.}~\bibnamefont
  {De~Lorenzo}}\ and\ \bibinfo {author} {\bibfnamefont {A.}~\bibnamefont
  {Perez}},\ }\href@noop {} {\bibfield  {journal} {\bibinfo  {journal}
  {Physical Review D}\ }\textbf {\bibinfo {volume} {97}},\ \bibinfo {pages}
  {044052} (\bibinfo {year} {2018})}\BibitemShut {NoStop}%
\bibitem [{\citenamefont {De~Lorenzo}\ and\ \citenamefont
  {Perez}(2019)}]{de2019light}%
  \BibitemOpen
  \bibfield  {author} {\bibinfo {author} {\bibfnamefont {T.}~\bibnamefont
  {De~Lorenzo}}\ and\ \bibinfo {author} {\bibfnamefont {A.}~\bibnamefont
  {Perez}},\ }\href@noop {} {\bibfield  {journal} {\bibinfo  {journal}
  {Physical Review D}\ }\textbf {\bibinfo {volume} {99}},\ \bibinfo {pages}
  {065009} (\bibinfo {year} {2019})}\BibitemShut {NoStop}%
\bibitem [{\citenamefont {Tian}(2005)}]{tian2005sitter}%
  \BibitemOpen
  \bibfield  {author} {\bibinfo {author} {\bibfnamefont {Y.}~\bibnamefont
  {Tian}},\ }\href@noop {} {\bibfield  {journal} {\bibinfo  {journal} {Journal
  of High Energy Physics}\ }\textbf {\bibinfo {volume} {2005}},\ \bibinfo
  {pages} {045} (\bibinfo {year} {2005})}\BibitemShut {NoStop}%
\bibitem [{\citenamefont {Good}\ \emph {et~al.}(2020)\citenamefont {Good},
  \citenamefont {Zhakenuly},\ and\ \citenamefont {Linder}}]{good2020mirror}%
  \BibitemOpen
  \bibfield  {author} {\bibinfo {author} {\bibfnamefont {M.~R.}\ \bibnamefont
  {Good}}, \bibinfo {author} {\bibfnamefont {A.}~\bibnamefont {Zhakenuly}}, \
  and\ \bibinfo {author} {\bibfnamefont {E.~V.}\ \bibnamefont {Linder}},\
  }\href@noop {} {\bibfield  {journal} {\bibinfo  {journal} {arXiv preprint
  arXiv:2005.03850}\ } (\bibinfo {year} {2020})}\BibitemShut {NoStop}%
\bibitem [{\citenamefont {Abramowitz}\ and\ \citenamefont
  {Stegun}(1999)}]{abramowitz1999ia}%
  \BibitemOpen
  \bibfield  {author} {\bibinfo {author} {\bibfnamefont {M.~S.}\ \bibnamefont
  {Abramowitz}}\ and\ \bibinfo {author} {\bibfnamefont {I.}~\bibnamefont
  {Stegun}},\ }\href@noop {} {\bibfield  {journal} {\bibinfo  {journal}
  {Washington: National Bureau of Standards}\ ,\ \bibinfo {pages} {923}}
  (\bibinfo {year} {1999})}\BibitemShut {NoStop}%
\bibitem [{\citenamefont {Rohde}\ \emph {et~al.}(2007)\citenamefont {Rohde},
  \citenamefont {Mauerer},\ and\ \citenamefont
  {Silberhorn}}]{rohde2007spectral}%
  \BibitemOpen
  \bibfield  {author} {\bibinfo {author} {\bibfnamefont {P.~P.}\ \bibnamefont
  {Rohde}}, \bibinfo {author} {\bibfnamefont {W.}~\bibnamefont {Mauerer}}, \
  and\ \bibinfo {author} {\bibfnamefont {C.}~\bibnamefont {Silberhorn}},\
  }\href@noop {} {\bibfield  {journal} {\bibinfo  {journal} {New Journal of
  Physics}\ }\textbf {\bibinfo {volume} {9}},\ \bibinfo {pages} {91} (\bibinfo
  {year} {2007})}\BibitemShut {NoStop}%
\bibitem [{\citenamefont {Weedbrook}\ \emph {et~al.}(2012)\citenamefont
  {Weedbrook}, \citenamefont {Pirandola}, \citenamefont
  {Garc{\'\i}a-Patr{\'o}n}, \citenamefont {Cerf}, \citenamefont {Ralph},
  \citenamefont {Shapiro},\ and\ \citenamefont
  {Lloyd}}]{weedbrook2012gaussian}%
  \BibitemOpen
  \bibfield  {author} {\bibinfo {author} {\bibfnamefont {C.}~\bibnamefont
  {Weedbrook}}, \bibinfo {author} {\bibfnamefont {S.}~\bibnamefont
  {Pirandola}}, \bibinfo {author} {\bibfnamefont {R.}~\bibnamefont
  {Garc{\'\i}a-Patr{\'o}n}}, \bibinfo {author} {\bibfnamefont {N.~J.}\
  \bibnamefont {Cerf}}, \bibinfo {author} {\bibfnamefont {T.~C.}\ \bibnamefont
  {Ralph}}, \bibinfo {author} {\bibfnamefont {J.~H.}\ \bibnamefont {Shapiro}},
  \ and\ \bibinfo {author} {\bibfnamefont {S.}~\bibnamefont {Lloyd}},\
  }\href@noop {} {\bibfield  {journal} {\bibinfo  {journal} {Reviews of Modern
  Physics}\ }\textbf {\bibinfo {volume} {84}},\ \bibinfo {pages} {621}
  (\bibinfo {year} {2012})}\BibitemShut {NoStop}%
\bibitem [{\citenamefont {Davies}\ and\ \citenamefont
  {Fulling}(1977)}]{davies1977radiation}%
  \BibitemOpen
  \bibfield  {author} {\bibinfo {author} {\bibfnamefont {P.~C.}\ \bibnamefont
  {Davies}}\ and\ \bibinfo {author} {\bibfnamefont {S.~A.}\ \bibnamefont
  {Fulling}},\ }\href@noop {} {\bibfield  {journal} {\bibinfo  {journal}
  {Proceedings of the Royal Society of London. A. Mathematical and Physical
  Sciences}\ }\textbf {\bibinfo {volume} {356}},\ \bibinfo {pages} {237}
  (\bibinfo {year} {1977})}\BibitemShut {NoStop}%
\bibitem [{\citenamefont {Tserkis}\ and\ \citenamefont
  {Ralph}(2017)}]{tserkis2017quantifying}%
  \BibitemOpen
  \bibfield  {author} {\bibinfo {author} {\bibfnamefont {S.}~\bibnamefont
  {Tserkis}}\ and\ \bibinfo {author} {\bibfnamefont {T.~C.}\ \bibnamefont
  {Ralph}},\ }\href@noop {} {\bibfield  {journal} {\bibinfo  {journal}
  {Physical Review A}\ }\textbf {\bibinfo {volume} {96}},\ \bibinfo {pages}
  {062338} (\bibinfo {year} {2017})}\BibitemShut {NoStop}%
\bibitem [{\citenamefont {Tserkis}\ \emph {et~al.}(2019)\citenamefont
  {Tserkis}, \citenamefont {Onoe},\ and\ \citenamefont
  {Ralph}}]{tserkis2019quantifying}%
  \BibitemOpen
  \bibfield  {author} {\bibinfo {author} {\bibfnamefont {S.}~\bibnamefont
  {Tserkis}}, \bibinfo {author} {\bibfnamefont {S.}~\bibnamefont {Onoe}}, \
  and\ \bibinfo {author} {\bibfnamefont {T.~C.}\ \bibnamefont {Ralph}},\
  }\href@noop {} {\bibfield  {journal} {\bibinfo  {journal} {Physical Review
  A}\ }\textbf {\bibinfo {volume} {99}},\ \bibinfo {pages} {052337} (\bibinfo
  {year} {2019})}\BibitemShut {NoStop}%
\bibitem [{\citenamefont {Bennett}\ \emph {et~al.}(1996)\citenamefont
  {Bennett}, \citenamefont {DiVincenzo}, \citenamefont {Smolin},\ and\
  \citenamefont {Wootters}}]{bennett1996mixed}%
  \BibitemOpen
  \bibfield  {author} {\bibinfo {author} {\bibfnamefont {C.~H.}\ \bibnamefont
  {Bennett}}, \bibinfo {author} {\bibfnamefont {D.~P.}\ \bibnamefont
  {DiVincenzo}}, \bibinfo {author} {\bibfnamefont {J.~A.}\ \bibnamefont
  {Smolin}}, \ and\ \bibinfo {author} {\bibfnamefont {W.~K.}\ \bibnamefont
  {Wootters}},\ }\href@noop {} {\bibfield  {journal} {\bibinfo  {journal}
  {Physical Review A}\ }\textbf {\bibinfo {volume} {54}},\ \bibinfo {pages}
  {3824} (\bibinfo {year} {1996})}\BibitemShut {NoStop}%
\bibitem [{\citenamefont {Wolf}\ \emph {et~al.}(2004)\citenamefont {Wolf},
  \citenamefont {Giedke}, \citenamefont {Kr{\"u}ger}, \citenamefont {Werner},\
  and\ \citenamefont {Cirac}}]{wolf2004gaussian}%
  \BibitemOpen
  \bibfield  {author} {\bibinfo {author} {\bibfnamefont {M.~M.}\ \bibnamefont
  {Wolf}}, \bibinfo {author} {\bibfnamefont {G.}~\bibnamefont {Giedke}},
  \bibinfo {author} {\bibfnamefont {O.}~\bibnamefont {Kr{\"u}ger}}, \bibinfo
  {author} {\bibfnamefont {R.}~\bibnamefont {Werner}}, \ and\ \bibinfo {author}
  {\bibfnamefont {J.~I.}\ \bibnamefont {Cirac}},\ }\href@noop {} {\bibfield
  {journal} {\bibinfo  {journal} {Physical Review A}\ }\textbf {\bibinfo
  {volume} {69}},\ \bibinfo {pages} {052320} (\bibinfo {year}
  {2004})}\BibitemShut {NoStop}%
\bibitem [{\citenamefont {Ivan}\ and\ \citenamefont
  {Simon}(2008)}]{ivan2008entanglement}%
  \BibitemOpen
  \bibfield  {author} {\bibinfo {author} {\bibfnamefont {J.~S.}\ \bibnamefont
  {Ivan}}\ and\ \bibinfo {author} {\bibfnamefont {R.}~\bibnamefont {Simon}},\
  }\href@noop {} {\bibfield  {journal} {\bibinfo  {journal} {arXiv preprint
  arXiv:0808.1658}\ } (\bibinfo {year} {2008})}\BibitemShut {NoStop}%
\bibitem [{\citenamefont {Marian}\ and\ \citenamefont
  {Marian}(2008)}]{marian2008entanglement}%
  \BibitemOpen
  \bibfield  {author} {\bibinfo {author} {\bibfnamefont {P.}~\bibnamefont
  {Marian}}\ and\ \bibinfo {author} {\bibfnamefont {T.~A.}\ \bibnamefont
  {Marian}},\ }\href@noop {} {\bibfield  {journal} {\bibinfo  {journal}
  {Physical review letters}\ }\textbf {\bibinfo {volume} {101}},\ \bibinfo
  {pages} {220403} (\bibinfo {year} {2008})}\BibitemShut {NoStop}%
\bibitem [{\citenamefont {Holevo}\ \emph {et~al.}(1999)\citenamefont {Holevo},
  \citenamefont {Sohma},\ and\ \citenamefont {Hirota}}]{holevo1999capacity}%
  \BibitemOpen
  \bibfield  {author} {\bibinfo {author} {\bibfnamefont {A.~S.}\ \bibnamefont
  {Holevo}}, \bibinfo {author} {\bibfnamefont {M.}~\bibnamefont {Sohma}}, \
  and\ \bibinfo {author} {\bibfnamefont {O.}~\bibnamefont {Hirota}},\
  }\href@noop {} {\bibfield  {journal} {\bibinfo  {journal} {Physical Review
  A}\ }\textbf {\bibinfo {volume} {59}},\ \bibinfo {pages} {1820} (\bibinfo
  {year} {1999})}\BibitemShut {NoStop}%
\bibitem [{\citenamefont {Onoe}\ \emph {et~al.}(2018)\citenamefont {Onoe},
  \citenamefont {Su},\ and\ \citenamefont {Ralph}}]{onoe2018particle}%
  \BibitemOpen
  \bibfield  {author} {\bibinfo {author} {\bibfnamefont {S.}~\bibnamefont
  {Onoe}}, \bibinfo {author} {\bibfnamefont {D.}~\bibnamefont {Su}}, \ and\
  \bibinfo {author} {\bibfnamefont {T.~C.}\ \bibnamefont {Ralph}},\ }\href@noop
  {} {\bibfield  {journal} {\bibinfo  {journal} {Physical Review D}\ }\textbf
  {\bibinfo {volume} {98}},\ \bibinfo {pages} {036011} (\bibinfo {year}
  {2018})}\BibitemShut {NoStop}%
\bibitem [{\citenamefont {Onoe}\ and\ \citenamefont
  {Ralph}(2019)}]{onoe2019universal}%
  \BibitemOpen
  \bibfield  {author} {\bibinfo {author} {\bibfnamefont {S.}~\bibnamefont
  {Onoe}}\ and\ \bibinfo {author} {\bibfnamefont {T.~C.}\ \bibnamefont
  {Ralph}},\ }\href@noop {} {\bibfield  {journal} {\bibinfo  {journal}
  {Physical Review D}\ }\textbf {\bibinfo {volume} {99}},\ \bibinfo {pages}
  {116001} (\bibinfo {year} {2019})}\BibitemShut {NoStop}%
\bibitem [{\citenamefont {Reznik}\ \emph {et~al.}(2005)\citenamefont {Reznik},
  \citenamefont {Retzker},\ and\ \citenamefont {Silman}}]{reznik2005violating}%
  \BibitemOpen
  \bibfield  {author} {\bibinfo {author} {\bibfnamefont {B.}~\bibnamefont
  {Reznik}}, \bibinfo {author} {\bibfnamefont {A.}~\bibnamefont {Retzker}}, \
  and\ \bibinfo {author} {\bibfnamefont {J.}~\bibnamefont {Silman}},\
  }\href@noop {} {\bibfield  {journal} {\bibinfo  {journal} {Physical Review
  A}\ }\textbf {\bibinfo {volume} {71}},\ \bibinfo {pages} {042104} (\bibinfo
  {year} {2005})}\BibitemShut {NoStop}%
\bibitem [{\citenamefont {Salton}\ \emph {et~al.}(2015)\citenamefont {Salton},
  \citenamefont {Mann},\ and\ \citenamefont
  {Menicucci}}]{salton2015acceleration}%
  \BibitemOpen
  \bibfield  {author} {\bibinfo {author} {\bibfnamefont {G.}~\bibnamefont
  {Salton}}, \bibinfo {author} {\bibfnamefont {R.~B.}\ \bibnamefont {Mann}}, \
  and\ \bibinfo {author} {\bibfnamefont {N.~C.}\ \bibnamefont {Menicucci}},\
  }\href@noop {} {\bibfield  {journal} {\bibinfo  {journal} {New Journal of
  Physics}\ }\textbf {\bibinfo {volume} {17}},\ \bibinfo {pages} {035001}
  (\bibinfo {year} {2015})}\BibitemShut {NoStop}%
\bibitem [{\citenamefont {Ver~Steeg}\ and\ \citenamefont
  {Menicucci}(2009)}]{ver2009entangling}%
  \BibitemOpen
  \bibfield  {author} {\bibinfo {author} {\bibfnamefont {G.}~\bibnamefont
  {Ver~Steeg}}\ and\ \bibinfo {author} {\bibfnamefont {N.~C.}\ \bibnamefont
  {Menicucci}},\ }\href@noop {} {\bibfield  {journal} {\bibinfo  {journal}
  {Physical Review D}\ }\textbf {\bibinfo {volume} {79}},\ \bibinfo {pages}
  {044027} (\bibinfo {year} {2009})}\BibitemShut {NoStop}%
\bibitem [{\citenamefont {Frolov}\ and\ \citenamefont
  {Serebriany}(1979)}]{frolov1979quantum}%
  \BibitemOpen
  \bibfield  {author} {\bibinfo {author} {\bibfnamefont {V.}~\bibnamefont
  {Frolov}}\ and\ \bibinfo {author} {\bibfnamefont {E.}~\bibnamefont
  {Serebriany}},\ }\href@noop {} {\bibfield  {journal} {\bibinfo  {journal}
  {Journal of Physics A: Mathematical and General}\ }\textbf {\bibinfo {volume}
  {12}},\ \bibinfo {pages} {2415} (\bibinfo {year} {1979})}\BibitemShut
  {NoStop}%
\bibitem [{\citenamefont {Obadia}\ and\ \citenamefont
  {Parentani}(2001)}]{obadia2001notes}%
  \BibitemOpen
  \bibfield  {author} {\bibinfo {author} {\bibfnamefont {N.}~\bibnamefont
  {Obadia}}\ and\ \bibinfo {author} {\bibfnamefont {R.}~\bibnamefont
  {Parentani}},\ }\href@noop {} {\bibfield  {journal} {\bibinfo  {journal}
  {Physical Review D}\ }\textbf {\bibinfo {volume} {64}},\ \bibinfo {pages}
  {044019} (\bibinfo {year} {2001})}\BibitemShut {NoStop}%
\bibitem [{\citenamefont {Obadia}\ and\ \citenamefont
  {Parentani}(2003)}]{obadia2003uniformly}%
  \BibitemOpen
  \bibfield  {author} {\bibinfo {author} {\bibfnamefont {N.}~\bibnamefont
  {Obadia}}\ and\ \bibinfo {author} {\bibfnamefont {R.}~\bibnamefont
  {Parentani}},\ }\href@noop {} {\bibfield  {journal} {\bibinfo  {journal}
  {Physical Review D}\ }\textbf {\bibinfo {volume} {67}},\ \bibinfo {pages}
  {024021} (\bibinfo {year} {2003})}\BibitemShut {NoStop}%
\bibitem [{\citenamefont {Brukner}(2014)}]{brukner2014quantum}%
  \BibitemOpen
  \bibfield  {author} {\bibinfo {author} {\bibfnamefont {{\v{C}}.}~\bibnamefont
  {Brukner}},\ }\href@noop {} {\bibfield  {journal} {\bibinfo  {journal}
  {Nature Physics}\ }\textbf {\bibinfo {volume} {10}},\ \bibinfo {pages} {259}
  (\bibinfo {year} {2014})}\BibitemShut {NoStop}%
\bibitem [{\citenamefont {Chiribella}\ \emph {et~al.}(2013)\citenamefont
  {Chiribella}, \citenamefont {D’Ariano}, \citenamefont {Perinotti},\ and\
  \citenamefont {Valiron}}]{chiribella2013quantum}%
  \BibitemOpen
  \bibfield  {author} {\bibinfo {author} {\bibfnamefont {G.}~\bibnamefont
  {Chiribella}}, \bibinfo {author} {\bibfnamefont {G.~M.}\ \bibnamefont
  {D’Ariano}}, \bibinfo {author} {\bibfnamefont {P.}~\bibnamefont
  {Perinotti}}, \ and\ \bibinfo {author} {\bibfnamefont {B.}~\bibnamefont
  {Valiron}},\ }\href@noop {} {\bibfield  {journal} {\bibinfo  {journal}
  {Physical Review A}\ }\textbf {\bibinfo {volume} {88}},\ \bibinfo {pages}
  {022318} (\bibinfo {year} {2013})}\BibitemShut {NoStop}%
\bibitem [{\citenamefont {Ara{\'u}jo}\ \emph {et~al.}(2014)\citenamefont
  {Ara{\'u}jo}, \citenamefont {Costa},\ and\ \citenamefont
  {Brukner}}]{araujo2014computational}%
  \BibitemOpen
  \bibfield  {author} {\bibinfo {author} {\bibfnamefont {M.}~\bibnamefont
  {Ara{\'u}jo}}, \bibinfo {author} {\bibfnamefont {F.}~\bibnamefont {Costa}}, \
  and\ \bibinfo {author} {\bibfnamefont {{\v{C}}.}~\bibnamefont {Brukner}},\
  }\href@noop {} {\bibfield  {journal} {\bibinfo  {journal} {Physical review
  letters}\ }\textbf {\bibinfo {volume} {113}},\ \bibinfo {pages} {250402}
  (\bibinfo {year} {2014})}\BibitemShut {NoStop}%
\bibitem [{\citenamefont {Branciard}\ \emph {et~al.}(2015)\citenamefont
  {Branciard}, \citenamefont {Ara{\'u}jo}, \citenamefont {Feix}, \citenamefont
  {Costa},\ and\ \citenamefont {Brukner}}]{branciard2015simplest}%
  \BibitemOpen
  \bibfield  {author} {\bibinfo {author} {\bibfnamefont {C.}~\bibnamefont
  {Branciard}}, \bibinfo {author} {\bibfnamefont {M.}~\bibnamefont
  {Ara{\'u}jo}}, \bibinfo {author} {\bibfnamefont {A.}~\bibnamefont {Feix}},
  \bibinfo {author} {\bibfnamefont {F.}~\bibnamefont {Costa}}, \ and\ \bibinfo
  {author} {\bibfnamefont {{\v{C}}.}~\bibnamefont {Brukner}},\ }\href@noop {}
  {\bibfield  {journal} {\bibinfo  {journal} {New Journal of Physics}\ }\textbf
  {\bibinfo {volume} {18}},\ \bibinfo {pages} {013008} (\bibinfo {year}
  {2015})}\BibitemShut {NoStop}%
\bibitem [{\citenamefont {Ho}\ \emph {et~al.}(2018)\citenamefont {Ho},
  \citenamefont {Costa}, \citenamefont {Giarmatzi},\ and\ \citenamefont
  {Ralph}}]{ho2018violation}%
  \BibitemOpen
  \bibfield  {author} {\bibinfo {author} {\bibfnamefont {C.}~\bibnamefont
  {Ho}}, \bibinfo {author} {\bibfnamefont {F.}~\bibnamefont {Costa}}, \bibinfo
  {author} {\bibfnamefont {C.}~\bibnamefont {Giarmatzi}}, \ and\ \bibinfo
  {author} {\bibfnamefont {T.~C.}\ \bibnamefont {Ralph}},\ }\href@noop {}
  {\bibfield  {journal} {\bibinfo  {journal} {arXiv preprint arXiv:1804.05498}\
  } (\bibinfo {year} {2018})}\BibitemShut {NoStop}%
\end{thebibliography}%

\end{document}